\RequirePackage{fix-cm}
\documentclass[smallextended]{svjour3}       % onecolumn (second format)
\smartqed  % flush right qed marks, e.g. at end of proof

\usepackage{amsmath, amssymb}%, wasysym}
\usepackage[english]{babel}
\usepackage[T1]{fontenc}
\usepackage{graphicx}
\usepackage{rotating}
\usepackage{array,multirow}
\newcommand{\R}{\mathbb R}

\def\be#1\ee{\begin{equation}#1\end{equation}}
\newcommand{\fer}[1]{(\ref{#1})}

\renewcommand{\thetable}{\arabic{section}.\arabic{table}}
\renewcommand{\thefigure}{\arabic{section}.\arabic{figure}}
\newcommand{\bq}{\begin{equation}}
\newcommand{\eq}{\end{equation}}
\newenvironment{equations}{\equation\aligned}{\endaligned\endequation}
\def\bqa{\begin{eqnarray}}
\def\eqa{\end{eqnarray}}

\def\e{\epsilon}

\newcommand{\bd}{\begin{displaymath}}
\newcommand{\ed}{\end{displaymath}}
\newcommand{\ba}{\begin{eqnarray}}
\newcommand{\ea}{\end{eqnarray}}

\def\R{\mathbb{R}}

 \journalname{
}
\begin{document}

\title{A multi-agent description of the influence of higher education on social stratification %\thanks{Grants or other notes
}
%\subtitle{Higher education and social stratification}

\titlerunning{Higher education and social stratification}        % if too long for running head

\author{Giacomo Dimarco       \and
        Giuseppe Toscani \and Mattia Zanella
}


\institute{G. Dimarco \at
              Department of Mathematics and Computer Science and Center for modeling, computing and statistics of the University of Ferrara, Italy. \\
              \email{ giacomo.dimarco@unife.it}           \\
           \and
           G. Toscani \at 
           Department of Mathematics "F. Casorati", University of Pavia, and IMATI CNR, Italy.  \\ \email{giuseppe.toscani@unipv.it}\\
              \and 
            M. Zanella \at Department of Mathematics "F. Casorati", University of Pavia, Italy.\\ \email{mattia.zanella@unipv.it}  
}

\date{Received: date / Accepted: date}

\maketitle

\begin{abstract}
We introduce and discuss a system of one-dimensional kinetic equations describing the influence of higher education in the social stratification of a multi-agent society. The system is obtained by coupling a model for knowledge formation with a kinetic description of the social climbing in which the parameters characterizing the elementary interactions leading to the formation of a social elite are assumed to depend on the degree of knowledge/education of the agents. In addition, we discuss the case in which the education level of an individual is function of the position occupied in the social ranking. With this last assumption we obtain a fully coupled model in which knowledge and social status influence each other. In the last part, we provide several numerical experiments highlighting the role of education in reducing social inequalities and in promoting social mobility. 
 
\end{abstract}

\section{Introduction}
\label{intro}

Educational expansion is usually regarded as a socially progressive development, able to reduce socioeconomic inequalities by providing a ladder of opportunities to a growing number of individuals. However, despite the evident increases in higher education enrolment rates, socioeconomic differences have so far shown no sign of equalizing \cite{Boli,Breen}. Indeed, while it is believed that higher education can serve as a vehicle for social mobility, the persistence of social inequalities in accessing higher education generally, and more prestigious forms of higher education in particular, seem to seriously compromise its potential effects. In reason of that, it has been recently outlined that higher education reproduces rather than reduce social inequality \cite{Marg} even if the debate is open \cite{Brown}.

On the other hand, while the existence of a social hierarchy is a diffuse concept which defies precise definitions, undoubtedly increasing advantages are associated with high socioeconomic status \cite{Dimaggio,Koski}. People in the highest part of the ladder, generally referred to as the upper class attend more prestigious schools, they are more influential in politics than people in the middle or working classes. High social status permits additionally to join elite social networks conferring benefits ranging from education to employment. It is also worth to mention that the social class is recognized as a strong social determinant of health \cite{Sa}.

To support the sociological analysis, and to quantify in a systematic way the possible relationships between education and social stratification, a substantial contribution can be furnished by mathematical modeling, resorting in particular to the methods and techniques of statistical mechanics. These are ideally suited to study social phenomena, which naturally include specific behavioral aspects of agents/individuals \cite{APZ,LMa,LMb,MD,PT13,PaVeZa,TTZ0,TTZ}. In economics, this approach attempted, among others, to justify the genesis of the formation of Pareto curves in wealth distribution of western countries \cite{ChaCha00,CCM,ChChSt05,CoPaTo05,DY00,GSV,SGD}, and to shed a light on the reasons behind opinion formation dynamics \cite{BN2,BN3,BN1,BeDe,Bou,Bou1,Bou2,CDT,DMPW,GGS,GM,Gal,GZ,SW,To1}.

Mathematical models dealing with the problem of social stratification have been introduced only recently, mainly by resorting to the dynamics of social networks in which individuals quest for high status in the social hierarchy \cite{BDL,Cha,KTessone,KTZ,ZTY}. An approach based on statistical mechanics has been introduced in \cite{DiTo}. In this work, throughout the well-established mathematical tools of kinetic theory of multi-agent systems \cite{PT13}, a microscopic mechanism leading to the formation of social stratification was introduced and discussed. In particular, it was possible to show that the consequent (explicit) macroscopic steady profile of the social hierarchy is characterized by a polynomial tail, heavily dependent on the parameters characterizing the microscopic interaction. Thus, the results in \cite{DiTo} well agree with the analysis of the economist Vilfredo Pareto \cite{Par1,Par2}, who, more than one century ago, observed that human societies tend to organize in a hierarchical manner, with the emergence of \emph{social elites}. Furthermore, Pareto noticed that social mobility in this hierarchical state appears to be higher in the middle classes than in the upper and lower part of the hierarchy, thus establishing a connection between the position occupied in the society with the role of higher education. 

Here, we start our investigation through the model of social stratification recently studied in \cite{DiTo} where the evolution in time of the density $f(w,t)$  of a system of agents characterized by a positive \emph{ranking value} $w$ was firstly introduced. The variation in time of the statistical distribution of the social rank has been obtained through the analysis of the elementary variations of the value $w$, classically defined as interactions in this context \cite{Cer,KT1,PT13}. In \cite{DiTo} the elementary upgrade of the ranking value has been modeled in the form
 \be\label{tru-coll}
 w_* = w  - \Psi^\e_\delta\left(\frac w {\bar w_L}\right) w + \eta_\e w,
 \ee
 where, for given positive constants $\e\ll 1$, $0<\mu<1$ and $0\le \delta \le 1$ 
  \be\label{vff}
 \Psi^\e_\delta(s) = - \mu \frac{e^{\e(s^{-\delta} -1)/\delta}-1}{(1-\mu)e^{\e(s^{-\delta} -1)/\delta}+1+\mu } , \quad  s \ge 0.
 \ee
In \fer{tru-coll} and \fer{vff}, the positive parameter $\e \ll1 $ quantifies the intensity of a single interaction while the role of the other parameters will be discussed later. The idea on which the model has been conceived is that the value $w$ of the social rank in \fer{tru-coll} can only be modified for two precise reasons, expressed by two different quantities. The first one is a quantity proportional to $w$ in which the coefficient $\Psi^\e_\delta(\cdot)$, which can assume assume both positive and negative values, is a function characterizing the asymmetric predictable behavior of agents acting in a society questing for a position in the social hierarchy. Within this modeling assumption each individual aims towards a desired target value $\bar w_L$, the desired social status. The second quantity, still proportional to $w$, takes into account a certain amount of unpredictability always present in human activities and that, for its nature, cannot be controlled. The random variable $\eta_\e$ describes these random variations, which in the mean are assumed negligible, and are characterized by a constant variance $\sigma\e$. Hence, the social status of individuals can be both increasing and decreasing by interactions with the background society, and the mean intensity of this variation is fully determined by the function $\Psi^\e_\delta$. While leaving the full description of the modeling reasons which lead to consider the elementary interaction in the form \fer{true-coll} to the next section, we only remark here that the positive parameter $\delta$ in \eqref{vff} is a measure of the difficulties of agents to climb the ladder starting from low values of the ranking $w$. This observation permits to naturally connect $\delta$ with a second model in which the education of each individual is properly described and supposed to influence the social status as discussed later.    

Given the elementary interaction \fer{tru-coll}, the time variation of the  density $f(w,t)$ obeys to a so-called linear
Boltzmann-like equation \cite{PT13,Vi}, fruitfully often written
in weak form. This weak form in fact corresponds to study the effect of the elementary interactions on observable smooth functions $\varphi(w)$ (the quantities of interest)
 \begin{equation}
  \label{kin-w}
 \frac{d}{dt}\int_{\R_+}\varphi(w)\,f(w,t)\,dx  = \frac 1{\e\tau} \mathbb E_{\eta_\e}
  \Big[ \int_{\R_+} \left(\frac w{u}\right)^\delta\, \bigl( \varphi(w_*)-\varphi(w) \bigr) f(w,t)
\,dw \Big].
 \end{equation}
  In \eqref{kin-w} $\mathbb E_{\eta_\e}[\cdot]$ denotes the expectation with respect to the random variable $\eta_\e$.  The constant quantity $u$ is the unit measure of social status, while the positive function $(w/u)^\delta$ is the frequency of the interactions of agents with social rank $w$. This choice for the frequency assigns a low pace to interactions to individuals with low rank, and assigns a high pace to interactions when the social status is greater. This assumption translates in a simple mathematical form that the motivations and the possibilities to climb the social ladder are stronger in individuals belonging to the middle and upper classes while individuals belonging to the lower class, having few possibility to succeed, are less tempted trying. Last, $\tau$ is a suitable relaxation time which simply identifies the time scale at which the formation of a social structure is reached.

Aiming to obtain the analytical expression of the possible steady-states of the above introduced model describing the formation of a social hierarchy, the analysis in \cite{DiTo} has been limited to fix in \fer{tru-coll} constant values for the parameters $\delta$ and $\sigma$. This choice corresponds to assign to all individuals in the system  the same possibilities to climb the social ladder. While this simplified assumption allows to characterize analytically the profile of the equilibrium state, a more realistic description of the social stratification was missing, since it would have required to take into account the natural inhomogeneity of the population, at least with respect to the education/knowledge level. 

In this work we move in this direction and we obtain a marked improvement in our modeling by resorting to a recent approach where the formation of collective knowledge was introduced \cite{PaTo} via a linear kinetic model. This approach allows to obtain the statistical distribution in time of the density $g(x,t)$ of individuals which at time $t \ge 0$ possess a given level of knowledge $x >0$. Identifying education with knowledge, one can take advantage from the kinetic model of knowledge formation by coupling it with the elementary interaction \fer{tru-coll}, in which the relevant parameters characterizing the social status $\delta$ and $\sigma$ are now assumed to depend on the knowledge variable $x$. 

In view of the meaning of the parameter $\delta$ to model the fact that individuals with higher knowledge, as expected \cite{Boli,Breen}, encounter less difficulties in increasing the social rank with respect to individuals with low knowledge, it is natural to introduce in \fer{vff} a function $\delta =\delta(x)$ decreasing with respect to the knowledge variable $x$. Also, by considering a variable variance $\sigma =\sigma(x)$ decreasing with respect to the knowledge variable $x$, one introduces into the model the property that individuals with higher knowledge suffer less risks in climbing and then reaching high social status with respect to individuals with low knowledge. An alternative approach would be to assign low variance also to individuals with very a low educational level, meaning that a low knowledge implies lower mobility chances. Additionally, we discuss the situations in which individuals belonging to the upper class have easier access to high quality institutions and consequently easily reach higher knowledge levels. This last aspect is introduced by considering that the amount of knowledge $x$ obtained by the background education system is a function of the social status $w$ of each agent. We remark that the above introduced choices of variable (with respect to the knowledge $x$ or with respect the social status $w$) parameters introduces a nonlinear coupling between the two unknowns, knowledge and social status, which does not allow to recover the distributions of social stratification and education in explicit form. Consequently, in a second part of this study, we rely on numerical experiments to put into light the properties of such model.  

The above described coupling will be studied in details in the rest of the paper. In more details, we will introduce the social climbing model obtained in \cite{DiTo} in the forthcoming Section \ref{model}. Then, the formation of collective knowledge, as considered in \cite{PaTo}, will be dealt with in Section \ref{knowledge}. The new model coming from the joint action of knowledge and social stratification will be introduced in Section \ref{model-full}. Last, numerical experiments will be presented in Section \ref{numerics} in which in particular the relation between inequalities and education will be put into light.

%%%%%%%%%%%%%%%%
\section{A kinetic model for social stratification}
\label{model}

Among other approaches, the description of social phenomena in a multi-agent system can be successfully obtained by resorting to statistical physics tools \cite{PT13}, and, in particular, through the methods borrowed from the kinetic theory of rarefied gases \cite{Vi}. The main goal on which this approach is based is to construct master equations of Boltzmann type, usually referred to as kinetic equations, describing the time-evolution of some characteristic of the agents, like wealth, opinion, knowledge, or, as in the case treated in this paper, of agent's ranking in the social ladder \cite{CaceresToscani2007,CCCC,DT,NPT,PT13,SC}.

The building block of the method is represented by the details of microscopic interactions, which, similarly to binary interactions between particles velocities in the classical kinetic theory, aim to describe the variation law of the selected agent's trait. Then, the microscopic law of variation of the number density consequent to the (fixed-in-time) way of interaction, is able to capture both the time evolution and the steady profile of the problem under consideration. 

In the case under study, the statistical distribution of the agents system is fully characterized by the unknown density $f = f(w, t)$ of the social rank $w$, $w \in \R_+$, occupied in the society by the agents at time $t\ge 0$. We assume that this value can be measured in terms of some reasonable unit $u$ which permits to translate in a mathematical form the concept of social hierarchy. Having in mind that there is a strong relationship between social rank and personal wealth, one possibility is to measure the value $w$ with the unit of the wealth of agents. Other choices can be equally possible.

The precise meaning of the density $f$ is the following. Given the system of individuals, and given an interval or a more complex sub-domain $D  \subseteq \R_+$, the integral
\[
\int_D f(w,t)\, dw
\]
represents the number of agents which are characterized by a rank $w \in D$ in the social ladder at time $t \ge 0$. It is assumed that the density function is normalized to one, that is for all $t\ge 0$
\be\label{uno}
\int_{\R_+} f(w,t)\, dw = 1.
\ee
This density $f(w,t)$ continuously changes in time as a consequence of elementary interactions with a background state: the society in which the individuals live and act. Similarly to the problems treated in \cite{DT,GT17,GT18}, the mechanism of social climbing in modern societies has been postulated in \cite{DiTo} to depend on some universal features that can be summarized by saying that agents likely tend to increase their status $w$ by interactions while manifest a certain resistance to decrease it, as theorized by Kahneman and Tversky in their seminal paper on decision under risk \cite{KT}.

An acceptable (from the sociological point of view) expression of the elementary variation of the social ranking of agents has been detailed in \cite{DiTo}, first identifying the \emph{mean values} which characterize, at least in a very stylized way, this multi-agent society. A first value, denoted by $\bar w$, identifies the upper limit of the low \emph{social} ranking. Below this value agents do not expect to be able to climb the social ladder. A second value, denoted by $\bar w_L$, with $\bar w_L > \bar w$, identifies the mean value that it is considered as the level of a satisfactory well-being by a large part of the population. %Note that both these values are in general differently perceived by individuals. 
The elementary interaction was consequently modeled to describe the behavior of agents in terms of these mean values in such a way to express the natural tendency of individuals to reach (at least) the value $\bar w_L$. However, in the agents behaviors, there is a strong asymmetry in the realization of this goal. %With respect to individuals which enjoy a high level of social ranking, 
This asymmetry expresses the objective difficulties to increase the value $w$, the social status, to reach the desired value $\bar w_L$ when very far from below. 

\begin{figure}\centering
    {\includegraphics[width=5.5cm]{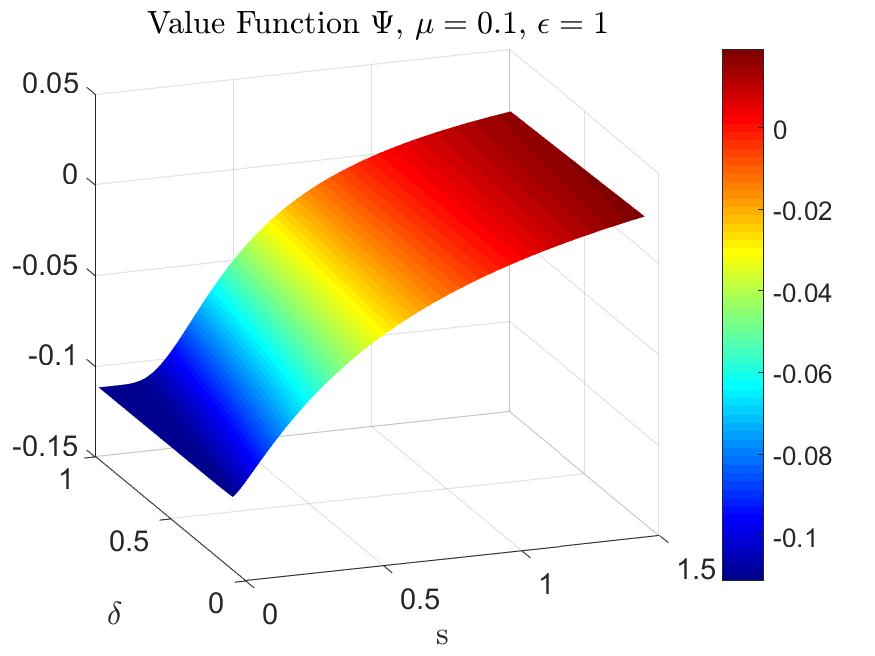}}
    \hspace{+0.35cm}
    {\includegraphics[width=5.5cm]{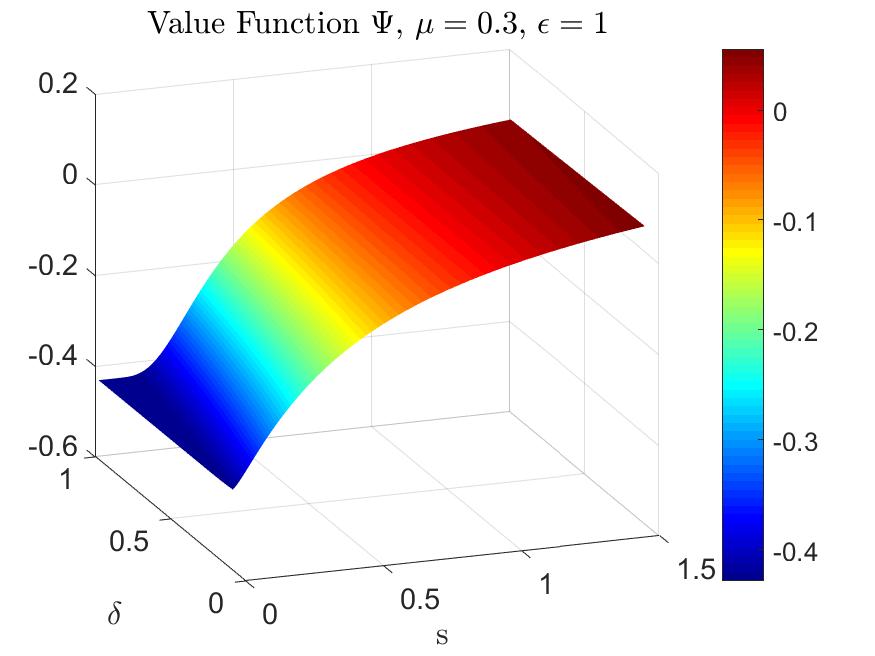}}\\
    \vspace{+0.45cm}
    {\includegraphics[width=5.5cm]{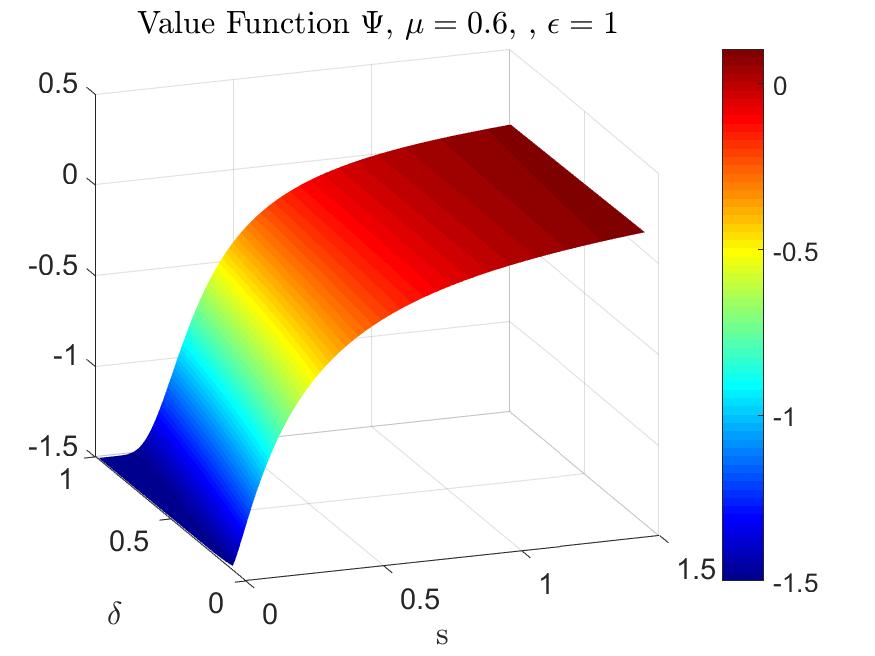}}
    \hspace{+0.35cm}
    {\includegraphics[width=5.5cm]{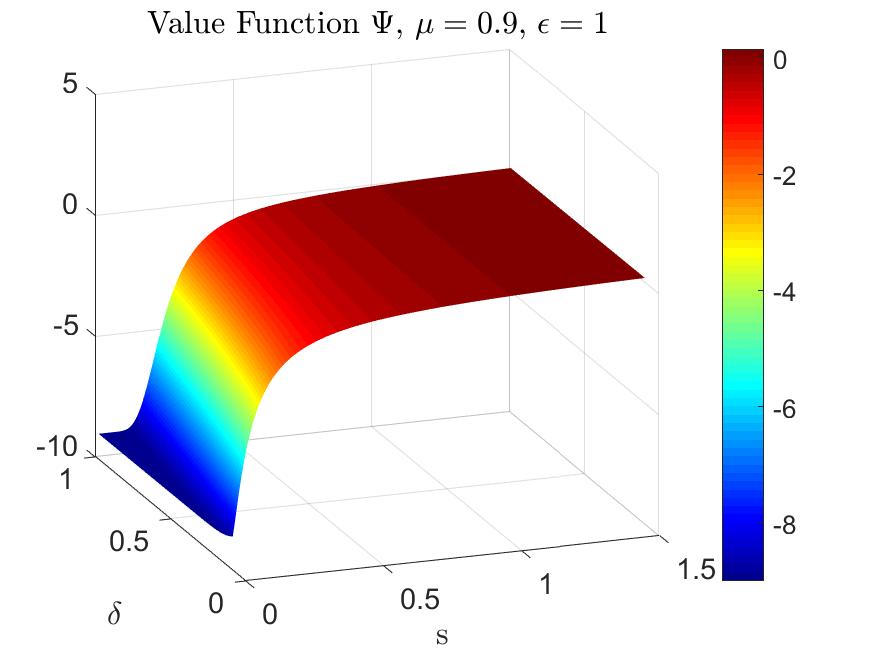}}\\
    \caption{Representation of the value function as a function of $s$ and $\delta$ for different values of $\mu$ and $\epsilon$.}\label{fig1}
\end{figure}

Taking into account the previous discussion and as briefly explained in the introduction, the elementary interaction has been modeled in the form
 \be\label{true-coll}
 w_* = w  - \Psi^\e_\delta\left(\frac w {\bar w_L}\right) w + \eta_\e w.
 \ee 
In agreement with the prospect theory originally proposed by Kahneman and Twersky in \cite{KT}, $\Psi^\e_\delta\left(s\right)$, as defined by \fer{vff}, plays the role of a so-called value function. This function is bounded from below and above, and satisfies the bounds
 \be\label{bbb}
- \frac\mu{1-\mu}  \le \Psi^\e_\delta(s) \le \mu \frac{1 - e^{-\e/\delta}}{(1-\mu)e^{-\e/\delta}+1+\mu }.
 \ee
Consequently, the positive constant $0<\mu<1$ characterizes the maximal amount of variation of the value function from above and below. Moreover,
starting from $s=0$, for any fixed value of the positive parameters $\e$ and $\delta$, the function $\Psi^\e_\delta(s)$ is convex in a small interval contained in the interval $(0,1)$, with an inflection point in $\bar s <1$, then concave. A picture of the value function for different values of the parameters is shown in Figure \ref{fig1}.   

The inflection point $\bar s <1 $ and the reference point $s =1$ are related to the previously mentioned mean values characterizing the climbing dynamics in the social ladder. The value $w = \bar w_L$, namely the perceived level of a satisfactory well-being corresponds to the reference point $s=1$. Instead, the upper limit of the low social ranking $\bar w$ characterizes the inflection point $\bar s <1$, in view of the relationship
 \be\label{www}
 \bar s = \frac{\bar w}{\bar w_L} < 1.
 \ee
Indeed, the graph of the function $\Psi^\e_\delta(s)$ can be split in the three regions $(0,\bar s)$, $(\bar s, 1)$ and $(1, +\infty)$, such that the graph is steeper in the middle region than in the other two, thus characterizing the middle region (the region of the middle social ranks) as the region in which it is present a higher value of the mobility, and consequently it is easier to improve the rank or to lose some position. On the contrary, the mobility is lower for values of the rank below $\bar w$ and above $\bar w_L$. The Figure \ref{fig1} clarifies also the role of the parameters $\mu$ and $\delta$. For large values of $\mu$ (bottom right image), the mobility is lower in the second and third part of the graph making more difficult for an agent to progress or descend into the social ladder. Concerning $\delta$ one can notice that the larger is this value, the larger is the region for which for individuals it is very difficult to climb the ladder. It is worth to remark that, as observed in \cite{DiTo}, the behavior of the function $\Psi^\e_\delta$ is in agreement with the original believe of Pareto \cite{Par2} that social mobility appears to be higher in the middle classes. It was also shown in \cite{DiTo} that the steady state of the underlying kinetic equation \fer{kin-w} characterizing the structure of the social ladder exhibits a polynomial decay at infinity, thus justifying the formation of a social elite as observed in the Western societies \cite{Scott}.

It is important to notice that the value of the inflection point $\bar w$ is increasing with $\delta$ enlarging the population belonging to the working class and the effects related to the difficulties to climb the social ladder. On the contrary, taking the limit $\delta \to 0$ in  \fer{vff} the value function converges to 
 \be\label{logn}
  \Psi^\e_0(s) = \mu \frac{s^\e -1}{(1+\nu)s^\e +1-\mu} , \quad  s \ge 0,
 \ee
namely to the value function considered in \cite{GT17,GT18}, leading to a lognormal steady profile. Note however that in this latter case the limit value function \fer{logn}, at difference with the value functions \fer{vff}, is concave, and the inflection point $\bar w$ is lost as it appears from Fig. \ref{fig1}.  

%Furthermore, in \eqref{true-coll} the term $\eta_\e$ is a zero-mean random variable with variance $\e\sigma$ taking into account possible deviations in the formation of the social rank. 

One of the main outcomes of the kinetic modeling of social stratification considered in \cite{DiTo} was related to the possibility to obtain an explicit expression of the steady distribution of the social rank. Following a well-established strategy of classical kinetic theory \cite{Vi}, subsequently generalized to cover kinetic equations for social and economic phenomena \cite{CoPaTo05,To1,GT-ec,PT13}, it is indeed possible to show that, for values of $\e \ll 1$ in \fer{tru-coll}  (limit of \emph{grazing interactions}), the solution of the kinetic equation \fer{kin-w} is close to the solution of the Fokker--Planck type equation
\[
\dfrac{\partial}{\partial t} f(w,t) = \dfrac{\partial}{\partial w} \left[ \dfrac{\mu}{2\delta} \left( 1-\left(\dfrac{\bar w_L}{w} \right)^\delta\right)w^{1+\delta}f(w,t) + \dfrac{\sigma }{2} \dfrac{\partial}{\partial w} \left( w^{2+\delta}f(w,t)\right)  \right]
\]
 for which the stationary solution is explicitly computable. This stationary solution is, in the case of social stratification, the function 
 \be\label{equilibrio}
f_\infty(w) =  f_\infty(\bar w_L) \left( \frac{\bar w_L}{w} \right)^{2 +\delta + \gamma/\delta} \exp\left\{ - \frac \gamma{\delta^2}\left( \left( \frac{\bar w_L}w \right)^\delta -1 \right)\right\},
 \ee 
where we denoted $\gamma = \mu/\sigma$. If one fixes the mass of the steady state \fer{equilibrio} equal to one, the consequent probability density is a particular case of the generalized Gamma distribution studied by Stacy \cite{Sta}. This is usually named Amoroso distribution \cite{Amo} and it is characterized by the polynomial tails identifying a class of people belonging to the upper class. The steady state depends on the typical parameters of the elementary interaction \fer{tru-coll}, namely $\delta$ and the quotient $\gamma = \mu/\sigma$. Clearly, for a given value of $\mu$, the values of $\delta$ and/or $\sigma$ are directly related to the size of the tail of the distribution see \eqref{equilibrio}. A small value of at least one of these parameters implies that the steady state distribution possesses a higher number of finite moments, which correspond to a less-marked social inequality. Note moreover that if the parameters $\delta$ and $\sigma$ are assumed simultaneously small, this effect is amplified by the presence of the product $\sigma\delta$.

\section{A kinetic model for distribution of education}\label{knowledge}

A kinetic description of knowledge formation was introduced in \cite{PaTo} with the aim of a better understanding the possible effects of knowledge in the distribution of wealth. Indeed, like it is supposed to happen in the case of social stratification treated here, different degrees of knowledge in a society are usually considered as one of the main causes of wealth inequalities \cite{inequalities}.

As for the case of social hierarchy, knowledge is a diffuse concept and in the following we refer to knowledge to as theoretical or practical understanding of a given subject acquired through experience or education such as information or skills. We will also often identify knowledge with education in the rest of the paper, even if we are aware of the limits of this identification. 

Let us now briefly explain the main motivations given in \cite{PaTo} to characterize the structure of the microscopic interactions which determine the individual knowledge. While knowledge is in part transmitted from the parents and family, the main factor that can enrich it, is the environment in which the individual grows and lives \cite{TB}. Indeed, the experiences that produce knowledge can not be fully inherited from the parents, such as the genome, but they are rather acquired over a lifetime. The learning process is a very complex phenomenon and it produces different results for each individual in a population. Even if all individuals are given the same opportunities, at the end of the cognitive process every individual appears to have a personal and different level of knowledge. Also, the personal knowledge is the result of a selection, which leads to retain mostly the notions that the individuals consider important, and to discard the rest.  

Given the above recalled facts and resorting once more to the legacy of kinetic theory, in \cite{PaTo} it was assumed that personal knowledge may be acquired as the result of a huge number of microscopic variations. Each microscopic variation is interpreted as cognitive process where a fraction of the knowledge is lost by virtue of selection, while at the same time the external background (the surrounding environment) furnishes a certain amount
of its knowledge to the individual. If one quantifies the nonnegative amount of knowledge of the individual with $x\in \R_+$, and with $z \in \R_+$ the knowledge achieved from the environment in a single process, the new amount of knowledge can be computed using the elementary upgrade
 \be\label{k1}
 x_* = (1-\e\lambda(x))x +\e \lambda_B(x) z + \psi_\e x.
 \ee
 As in \fer{tru-coll}, the positive parameter $\e \ll1 $ quantifies the intensity of a single cognitive process. In \fer{k1} the functions $\lambda(x)$ and $\lambda_B(x)$ are related, respectively, to the amounts of selection and external learning, while $\psi_\e$ is a random parameter of zero mean and variance $\e\nu$ which takes into account the possible unpredictable modifications of the knowledge process. The time evolution of the distribution of knowledge  $g(x,t)$ may be chosen to obey to a Boltzmann-like equation as for the case of the formation of a social structure. In this case, the study of the effect of the elementary interactions on observable smooth functions $\varphi(x)$ gives 
 \be
  \label{line-x}
  \begin{split}
  &\frac{d}{dt}\int_{\R_+}g(x,t)\varphi(x)\,dx \\
  &\quad = \frac 1{\tau\e}
  \mathbb E_{\psi_\e}\left[ \int_{\R_+^2} \bigl( \varphi(x_*)-\varphi(x) \bigr) g(x,t)M(z)
\,dx\,dz \right],
\end{split}
 \ee
where the post-interaction variable $x_*$ is given by \fer{k1}, and $M(z)$ denotes the statistical distribution of the background knowledge, that is assumed to possess a finite mean value $M_B$. Note that we assumed that the knowledge process is characterized by the same relaxation time of the social stratification process even if different choices are possible. In \eqref{line-x}, we denoted by $\mathbb E_{\psi_e}[\cdot]$ the expectation with respect to the random variable $\psi_\e$. 
The main result in \cite{PaTo} was to show that, if $\e \ll 1$ i.e. in the so-called \emph{grazing interactions} regime, similarly to the case of the social climbing model, the stationary solution of equation \fer{line-x} is characterized by a polynomial decay at infinity. This is a nice way to say that the model is in agreement with the observation of the existence in the society of a (very small) class of people possessing a very high knowledge level or, recast in other words, a class of people with an extraordinary intellectual or creative power exists in our society that stand out from the rest. As discussed in Section \ref{model}, if $\e\ll 1$, one can further show that, whenever the functions $\lambda(x)$ and $\lambda_E(x)$ characterizing the elementary interaction in the process of knowledge formation are assumed constant, the stationary solution of the kinetic equation \fer{line-x} is close to the solution of a Fokker--Planck type equation for which the steady state is explicitly computable. This stationary solution is the inverse gamma density function 
 \be\label{equilibrio2}
g_\infty(x) =  \frac{\theta^\kappa}{\Gamma(\kappa)} \frac 1{x^{\kappa +1}}\exp\left\{-\frac\theta{x} \right\},
 \ee 
where 
\[
\theta = \frac{\lambda_BM_B}\nu; \quad \kappa = 1 + \frac \lambda\nu.
\]
Note that the mean value of the steady state is finite
\be\label{mean}
\int_{\R_+} x g_\infty(x) \, dx = \frac{\lambda_B}\lambda M_B,
\ee
and it is measured by the quotient $\lambda_B/\lambda$ of the constant parameters of external learning and selection characterizing the elementary interaction \fer{k1}. 

It is important to remark that, in this model, the size $\kappa$ of the tail of the steady solution \fer{equilibrio2} does not depend on the amount of external learning $\lambda_B$, but only on the quotient between the selection parameter $\lambda$, and  the variance $\nu$ of the random part of the interaction. 

\begin{figure}[ht!]
   \includegraphics[width=\textwidth]{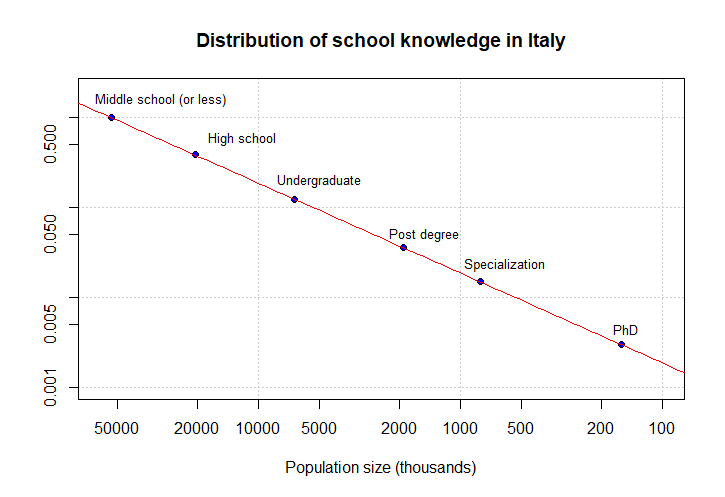}
   \centering
   \caption{Distribution of school knowledge in Italy by highest degree, 2011 census.\label{fig:knowledge}}
   \label{fig:know}
\end{figure}

\paragraph{Distribution of scholar education.} 
In \cite{GT-ec} the capability of the model to reproduce existing distributions originating from different social phenomena and in particular the behavior of the tails of the steady state of equation \fer{line-x} has been tested by looking at the distribution of scholar education in Italy, using data collected from the 2011 census (cfr. Figure \ref{fig:knowledge}). 
This plot is based on the data given in Table \ref{tab:knowledge}, which collects the number of citizens per type of school degree in the second column, and the (inverse) cumulated number of people for school degree in the fourth column.
The first basic level of school knowledge includes every citizen who holds the middle school degree as highest degree, which corresponds nowadays to the minimum Italian compulsory education level.
The second and third levels include people who got a high school degree and an undergraduate degree as highest degree, respectively. 
The fourth level includes the $1\,124\,802$ Italians who got a ``short'' (less than 1 year of studies) post graduate degree. 
Finally, the last two levels give the number of citizens holding either a ``specialization'' or PhD as highest school degree, which are respectively 634 503 and 159 455 citizens, a very small percentage of the whole Italian population. 

Figure \ref{fig:knowledge}, expressed in $\log-\log$ scale, clearly shows that this empirical (inverse) cumulated distribution exhibits a tail which can be put in relation with the steady states \eqref{line-x} obtained in \cite{PaTo}.

\begin{table}[ht!]
\begin{center}
\begin{tabular}{ l  r  r  r r }
{\bf Highest degree} \hspace{0.5cm} & {\bf Citizens} & {\bf (\%)} & {\bf Cumulated values} & {\bf (\%)}\\ \hline
Middle school & 32 906 278 & 61.6\% & 53 422 830 & 100.0\% \\
High school & 13 906 688 & 26.0\% & 20 516 552 & 38.4\% \\
Undergraduate &4 691 104 & 8.8\% & 6 609 864 & 12.4\% \\
Post degree &1 124 802 & 2.1\% &  1 918 760 & 3.6\% \\
Specialization &634 503 & 1.2\% & 793 958 & 1.5\% \\
PhD &159 455 & 0.3\% & 159 455 & 0.3\% \\ \hline
(totals) & 53 422 830 & 100.0\% &
\end{tabular}
\caption{Tabular data of the distribution of school knowledge in Italy, 2011 census.\label{tab:knowledge}}
\end{center}
\end{table}

\section{A kinetic model joining education and social stratification}\label{model-full}
\setcounter{equation}{0}

In this section, we will merge the model for social climbing \fer{kin-w} whose details have been given in Section \ref{model}, with the model for knowledge formation \fer{line-x} detailed in Section \ref{knowledge}. In its original formulation, the social climbing interaction was described in terms of the two constant parameters $\delta,\sigma$.  Indeed, as shown in \cite{DiTo}, and briefly recalled in Section \ref{model}, the parameters $\delta$ and $\sigma$ fully characterize the steady state of the social stratification. Suppose now that these parameters depend on the personal knowledge of the agent. One reasonable assumption would be that an individual uses the personal knowledge to improve the possibilities to climb the social ladder. This effect can be obtained by assuming that the personal knowledge modifies the value of the parameter $\delta$. Since this parameter is directly related to the length of the interval in which the value function is convex and consequently, in this interval, it is difficult to increase the value of the rank $w$. Thus, one reasonable choice is to fix the function $\delta(x)$ as a decreasing function of $x$, which reflects the idea that personal knowledge could be fruitfully employed to reduce the interval of convexity of the value function. A possible choice is
\be\label{alfa}
 \delta(x) = \frac {\delta_0}{(1+x)^{\alpha}},
 \ee
 with $\alpha >0$ and $\delta_0$ a suitable constant. Also, it can be reasonably assumed that the random part of the elementary interaction \fer{tru-coll}, characterized by the value of the variance $\sigma$ decreases as $x$ increases, thus indicating that knowledge helps in reducing the randomness present in the social climbing process. As before, we can assume that the variance of the random variable $\eta_\e$ is such that 
\be\label{beta}
 \sigma(x) = \frac {\sigma_0}{(1+x)^{\beta}},
 \ee
 with $\beta >0$ and $\sigma >0$. An alternative to \eqref{beta} consists in considering the following law for the variance of $\eta_\e$
 \be\label{beta1}
 \sigma(x) = \frac{m_{\sigma}+M_{\sigma}x^{-\chi}}{1 + x^{-\chi}}
 \ee
 with $\chi>0$ again. This second possibility furnishes a slightly different variance function which ranges from a minimum $m_{\sigma}$ to a maximum value $M_{\sigma}$ and it is such that as before the random possibility to descend towards lower social positions is reduced as the knowledge increases. However, this new function \eqref{beta1} imposes a limit to a possible minimum value of the variance even for individuals with a larger education level. Moreover, independently on the value of $\chi$ individuals with an average level of knowledge shares the same stochastic behavior in terms of elementary interactions compared to the case \eqref{beta} in which as $\beta$ increases the stochasticity in the formation of the social ladder is lower for individuals with the same knowledge level. A last remarkable example, we aim to introduce and discuss, consists in fixing the variance of the random process characterizing the social climbing dynamics in such a way that individuals having an average education level are the ones who more likely are subject to unpredictability and may descend or climb as a result of a stochastic event. In other words, we assume that it is unlikely that people with a very low education level has large chances to jump to higher social positions and at the same time that individuals with a high knowledge level cannot easily descend to lower positions in the ladder as a result of a random event. Let observe that this last choice is in agreement with the behavior of the value function $\Psi^\e_\delta$ that social mobility appears to be higher in the middle classes. For this last example, the function characterizing the variance reads as
 \be\label{beta2}
 \sigma(x) = \frac {\sigma_0}{1+(x-M_B)^{\beta^\prime}}.
 \ee

Once the above discussed coupling choices are done, the evolution of the joint action of knowledge and social rank in the system of agents can be described in terms of the density $f=f(x,w,t)$ of agents which at time $t >0$ are represented by their knowledge $x \in \R_+$ and social status $w\in \R_+$. Following the trail of kinetic theory, the evolution in time of the density $f$ is then described by the following equation (in weak form)
 \begin{equations}\label{kine-xw}
&\frac{d}{dt}\int_{\R_+^2}\varphi(x,w) f(x,w,t)\,dx\, dw  = \\
&\quad\frac 1{\tau\e}\mathbb E_{\eta_\e,\psi_\e} \left[\int_{\R_+^3}  \left(\frac w{u}\right)^\delta\,\bigl( \varphi(x_*,w_*) - \varphi(x,w) \bigr) f(x,w,t)M(z)
\,dx\, dw \right],
 \end{equations}
 being $\mathbb E_{\eta_\e,\psi_\e}[\cdot]$ the expectation with respect to the random variables $\eta_\e$ and $\psi_\e$.
In \fer{kine-xw} the post-interaction pair $(x_*, w_*)$  is given by 
\be\label{kn}
x_* = (1-\e\lambda)x + \e\lambda_B z + \psi_\e x, 
\ee
and
\be\label{so}
 w_* = w  - \Psi^\e_{\delta(x)}\left(\frac w {\bar w_L}\right) w + \eta_\e(x) w.
\ee
Note that within this choice, the elementary interaction characterizing the evolution of the collective knowledge is still assumed independent of the value $w$ of the social ranking. This assumption is reasonably acceptable for countries where school education is overwhelmingly public and the individual has the possibility of accessing almost any school, regardless of position on the social ladder. Even if not general, within this choice one can infer the precise influence of the education on the social stratification. Following this path, if the test function $\varphi$ is independent of $w$, that is $\varphi=\varphi(x)$, equation \fer{kine-xw} reduces to the equation \fer{line-x} for the marginal density of knowledge $g(x,t)$. 

Instead, a more general assumption on the pair \fer{kn}--\fer{so} implies that the amount of knowledge furnished by the background, namely a better possibility of education, could depend on the social status of the individual, in a way that a higher social position would in general correspond to a possibility to have higher and better education as experimentally observed. A realization of this assumption is obtained by multiplying in \fer{kn} the coefficient $\lambda_B$, quantifying the amount of external learning, by a suitable increasing function $E(w)$ of the social rank $w$ in the form
 \be\label{edu}
 E(w) = \frac{m+Mw^\xi}{1 + w^\xi},
 \ee
where $\xi, m, M$ are positive constants. The function in \fer{edu} ranges from a minimal level of education $m$ to a maximal one, indicated by $M>m$. Consequently, the elementary interaction \fer{kn} is substituted by 
\be\label{kn1}
x_* = (1-\e\lambda)x + \e\lambda_BE(w) z + \psi_\e x. 
\ee
The pair \fer{kn1}--\fer{so} will now produce a coherent evolution of the joint process for education and social rank in which each process is intimately dependent from the other. As far as the new evolution process for the knowledge is concerned, the mean value of the steady state relative to the elementary interaction \fer{kn1} is now given by formula \fer{mean}, which reads
 \be\label{mean2}
\int_{\R_+} x g_\infty(x) \, dx = \frac{\lambda_B}\lambda M_BE(w).
\ee
Thus, the effect of the social rank is reflected on the mean value of the education process by multiplying the mean \fer{mean} by the function $E(w)$. This induces a situation in which the average level of knowledge is larger for individuals with larger means.
Finally, it is worth to observe that, as remarked at the end of Section \ref{knowledge}, the dependence of knowledge from the social rank, resulting in the elementary interaction \fer{kn1}, for a given value of $w$ does not modify the size of the tail of the steady profile of the knowledge distribution.
 
\section{Numerical experiments}\label{numerics}
This section contains a numerical description of the kinetic model coupling education with the social stratification in a population where individuals quest for reaching a high status in the social hierarchy. The idea of the numerical experiments is to highlight the role played by education in the formation of Western societies in a hierarchical manner, with the emergence of social elites. At the same time and in a similar way, we numerically explore the relation between the possibility of reaching high level of education jointly with high social status, reflecting the evidence that individuals with larger means have an easier access to top ranked schools and Universities and consequently, as it is natural to suppose, to a larger level of knowledge.

The present section is divided into three parts in which we study different aspects of the model introduced in the Sections \ref{model}-\ref{knowledge} and in Section \ref{model-full}. In particular, we start by showing solutions to the Boltzmann-type equations in which interactions are defined by \eqref{kn} and \eqref{so} together with \eqref{alfa} and \eqref{beta} respectively for the knowledge and the social structure and we compare them to the case in which social hierarchy and education are independent phenomena. In a second test, we discuss the case in which knowledge depends additionally on the social status of individuals \eqref{kn1}. For all these cases, we explore solutions for two different regimes identified by the scaling parameter $\epsilon$. This choice permits in the regime $\epsilon\ll 1$ to obtain a good approximation of Fokker-Planck-type equations when social hierarchy and education are considered independent dynamics %. We recall that for independent dynamics, 
and the analytical equilibrium states are given by \eqref{equilibrio} and \eqref{equilibrio2}. Finally, in a last part of the section, we measure and discuss the possible inequalities arising both in the education as well as in the social stratification thanks to a thorough numerical exploration of different modeling choices for the coupling functions $\delta(x),\sigma(x),E(w)$. 

For the numerical approximation of the Boltzmann equation, we apply a Monte Carlo method as described for instance in \cite{PT13}. The simulations are run using $N= 10^6$ agents (particles). 

In what follows we will use the notations:
\begin{itemize}
 \item
 $X(t)$ to denote the random variable representing the amount of know\-ledge
 of the population at time $t>0$. Its density is given by the
 solution of equation \fer{line-x} in the decoupled case. Hence
  \[
  G(x, t) \,dx = P(X(t) \in (x, x+dx)), \quad x \ge 0.
  \]
We will denote as well by $\mathcal{G}$ the distribution function relative to $X(t)$
 \be\label{disg}
\mathcal{G}(x) = P(X(t) \leq x) = \int_{0}^{x} G(y,t) \, dy.
 \ee
 and the complementary cumulative distribution
\be\label{cumg}
\mathcal{\bar{G}}(x)=1-\mathcal{G}(x)
\ee
which will be useful to identify the tail behavior of the steady state.
\item
$W(t)$ to denote the random variable which represents the social status of the individuals in the population at time $t>0$. Given the solution of equation \fer{kine-xw}, its density is given by the marginal density
  \[
  F(w, t) \,d w= P(W(t) \in (w, w+dw))= \,d w\int_\R  f(x,w,t)\, dx.
  \]
We will denote as well by $\mathcal{F}$ the distribution function
\be\label{disf}
\mathcal{F}(w) = P(W(t) \leq w) = \int_{0}^{w} F(v,t) \, dv.
 \ee
  and the complementary cumulative distribution
 \be\label{cumf}
 \mathcal{\bar{F}}(w)=1-\mathcal{F}(w)
 \ee
which will be useful as for the case of knowledge to identify the slope of the tail of the equilibrium state and consequently the possible formation of social elites.
 
\end{itemize}

\subsection{Test 1}
In this first test, we consider the case of equations \eqref{kn}-\eqref{so}, i.e. the case in which the formation of a social hierarchy depends on the level of knowledge of each individual but not vice-versa. The function $\delta(x)$ giving the shape of the probability distribution describing the social structure is fixed in similar way of suggested in \eqref{alfa} while the variance of the random variable $\eta_\e = \e \sigma(x)$ modeling unpredictable deviations during the process of formation of the social hierarchy is fixed using \eqref{beta}. To be more precise we use the following relations 
\begin{equations}\nonumber
\displaystyle &\sigma(x) = \frac {\sigma_0}{(1+x)^{\beta}}, \quad \delta(x) = \frac {\delta_0}{(1+x)^{\alpha}},\\
 & \sigma_0=0.5, \quad \delta_0=0.1,\quad \alpha=2, \quad \beta=2.
\end{equations}
We also define $\mu=0.5$ and $\bar w_L=1$ for the social climbing model. We then take $\lambda = \lambda_B = \nu= 0.1$ and $M_B=1$ in the knowledge model. We finally consider a time step of $\Delta t = 0.1$, $\epsilon=0.1$ and a final computation time of $t =100$, where the steady state is numerically measured to be practically reached, i.e. for larger times the solution remains unchanged. With this choice of the parameters, we suppose the model to be in the so-called Boltzmann regime in contrast with the Fokker--Planck regime for which the parameters in the interactions laws will be conveniently scaled in order for the model to produce results in the grazing collision regime. 

%\begin{figure}[!htbp]
%   \begin{center}
%       \begin{tabular}{cc}  \hspace{-0.5cm}       
%               \includegraphics[width=0.5\textwidth]{figures/test1/test0_p}   \hspace{0.5cm}  
%           \includegraphics[width=0.5\textwidth]{figures/test1/test0} \\ \hspace{-0.5cm}
%           \includegraphics[width=0.5\textwidth]{figures/test1/test1_pbis}   \hspace{0.5cm}  
%           \includegraphics[width=0.5\textwidth]{figures/test1/test1_bis} 
%       \end{tabular}
%       \caption{Test 1: random subset of $N = 1000$ agents of the kinetic model (left) and the kinetic density (right). Top images show the decoupled case: $\sigma(x)=\sigma_0$ and $\delta(x)=\delta_0$. Bottom images show the case in which the social structure depends upon knowledge as defined in \eqref{alfa} and \eqref{beta}. }
%       \label{fig:test1}
%   \end{center}
%\end{figure}
\begin{figure}[!htbp]
    \begin{center}
        %\begin{tabular}{cc}  \hspace{-0.5cm}       
            \includegraphics[width=0.45\textwidth]{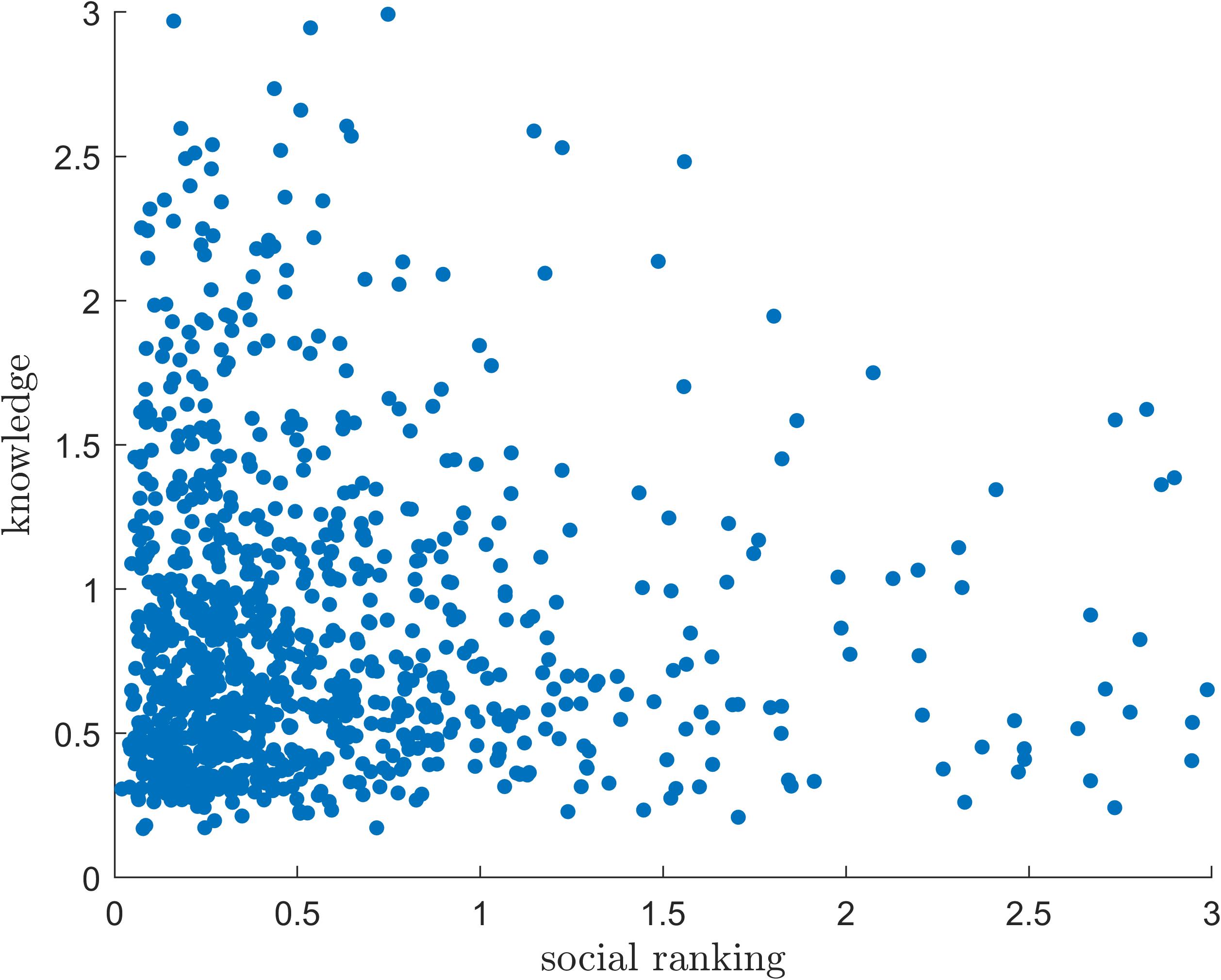}  % \hspace{0.5cm}  \hspace{-0.5cm}
            \includegraphics[width=0.45\textwidth]{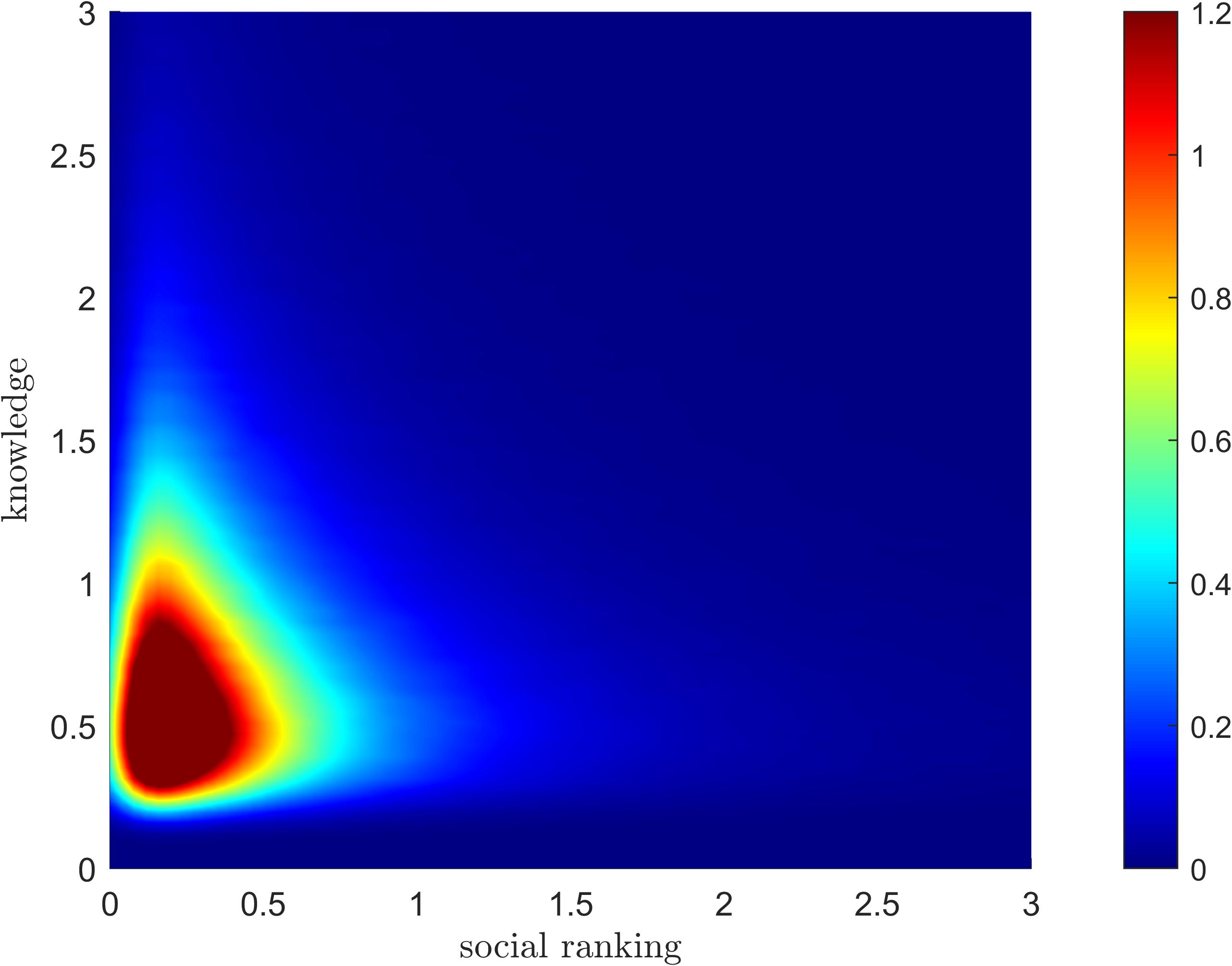} \\ \vspace{0.8cm}
            \includegraphics[width=0.45\textwidth]{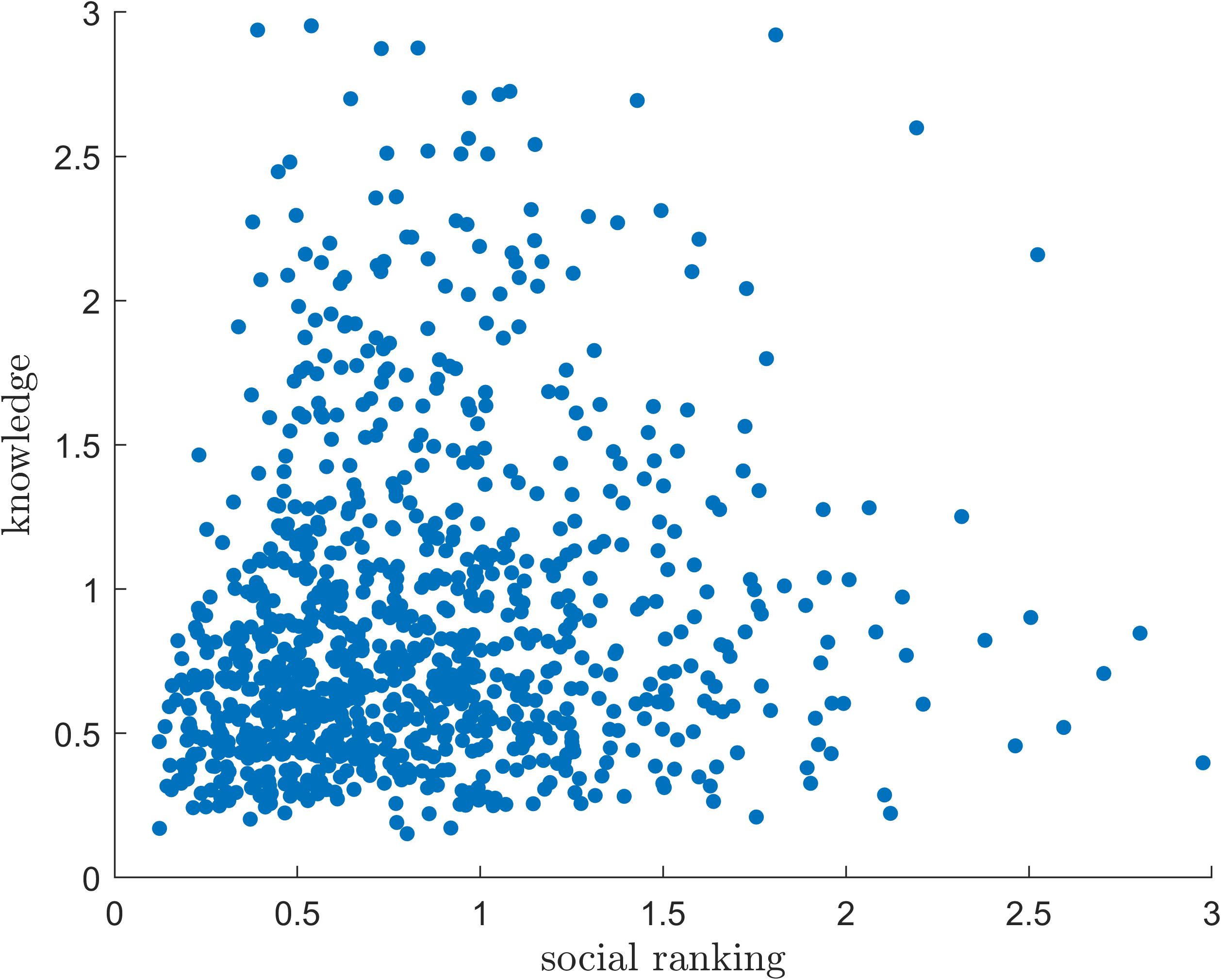}   %\hspace{0.5cm}  
            \includegraphics[width=0.45\textwidth]{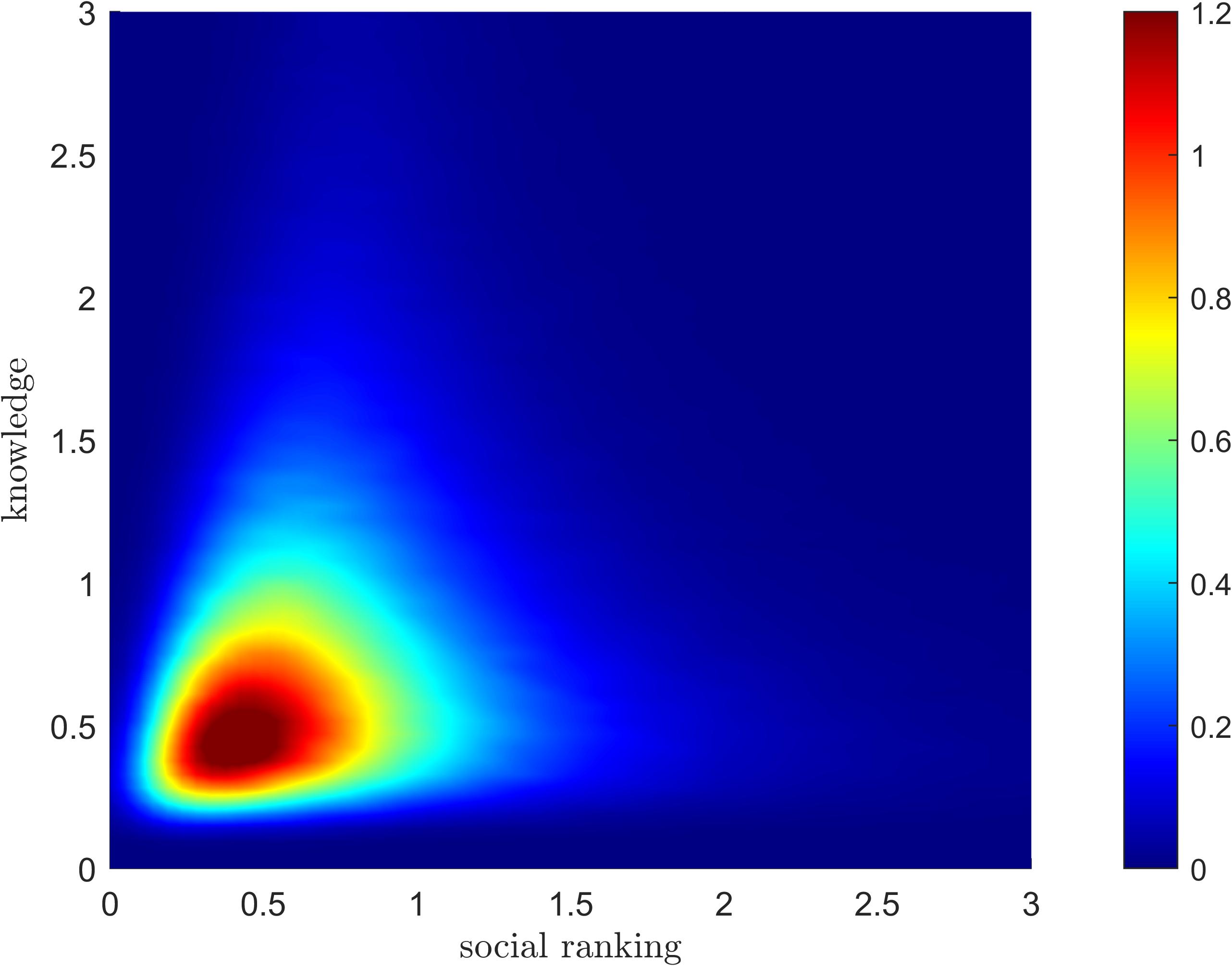} 
        %\end{tabular}
        \caption{Test 1: random subset of $N = 1000$ agents of the kinetic model (left) and the kinetic density (right). Top images show the decoupled case: $\sigma(x)=\sigma_0$ and $\delta(x)=\delta_0$. Bottom images show the case in which the social structure depends upon knowledge as defined in \eqref{alfa} and \eqref{beta}. }
        \label{fig:test1}
    \end{center}
\end{figure}

%\begin{figure}[!htbp]
%   \begin{center}
%       \begin{tabular}{cc}    \hspace{-0.5cm}
%           \includegraphics[width=0.5\textwidth]{figures/test1/test1_marginals_3}
%           \hspace{0.5cm}
%           \includegraphics[width=0.5\textwidth]{figures/test1/test1_marginals_tails_3}  \\  
%       \end{tabular}
%       \caption{Test 1. Left image shows the two marginal densities relative to knowledge and social status. Right image shows the complementary cumulative marginal distributions for the same quantities in log-log scale. The slopes $p$ of the tails are estimated from $\mathcal{\bar{F}}(w)\approx w^{-p}$ and $\mathcal{\bar{G}}(x)\approx x^{-p}$ for larger $w$ and $x$.
%       }
%       \label{fig:test2}
%   \end{center}
%\end{figure}

\begin{figure}[!htbp]
    \begin{center}
        %\begin{tabular}{cc}  \hspace{-0.5cm}
            \includegraphics[width=0.45\textwidth]{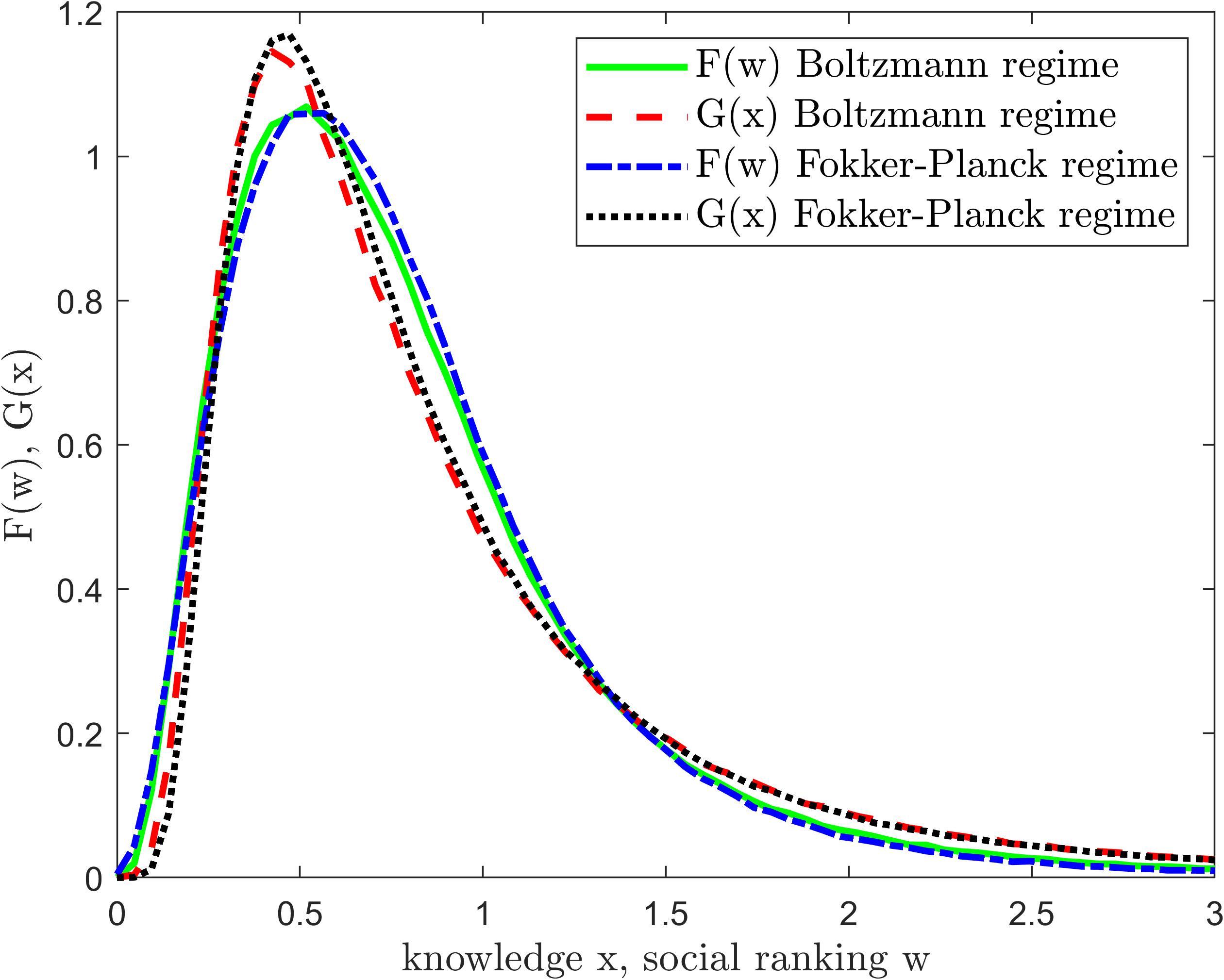}
            \hspace{0.7cm}
            \includegraphics[width=0.45\textwidth]{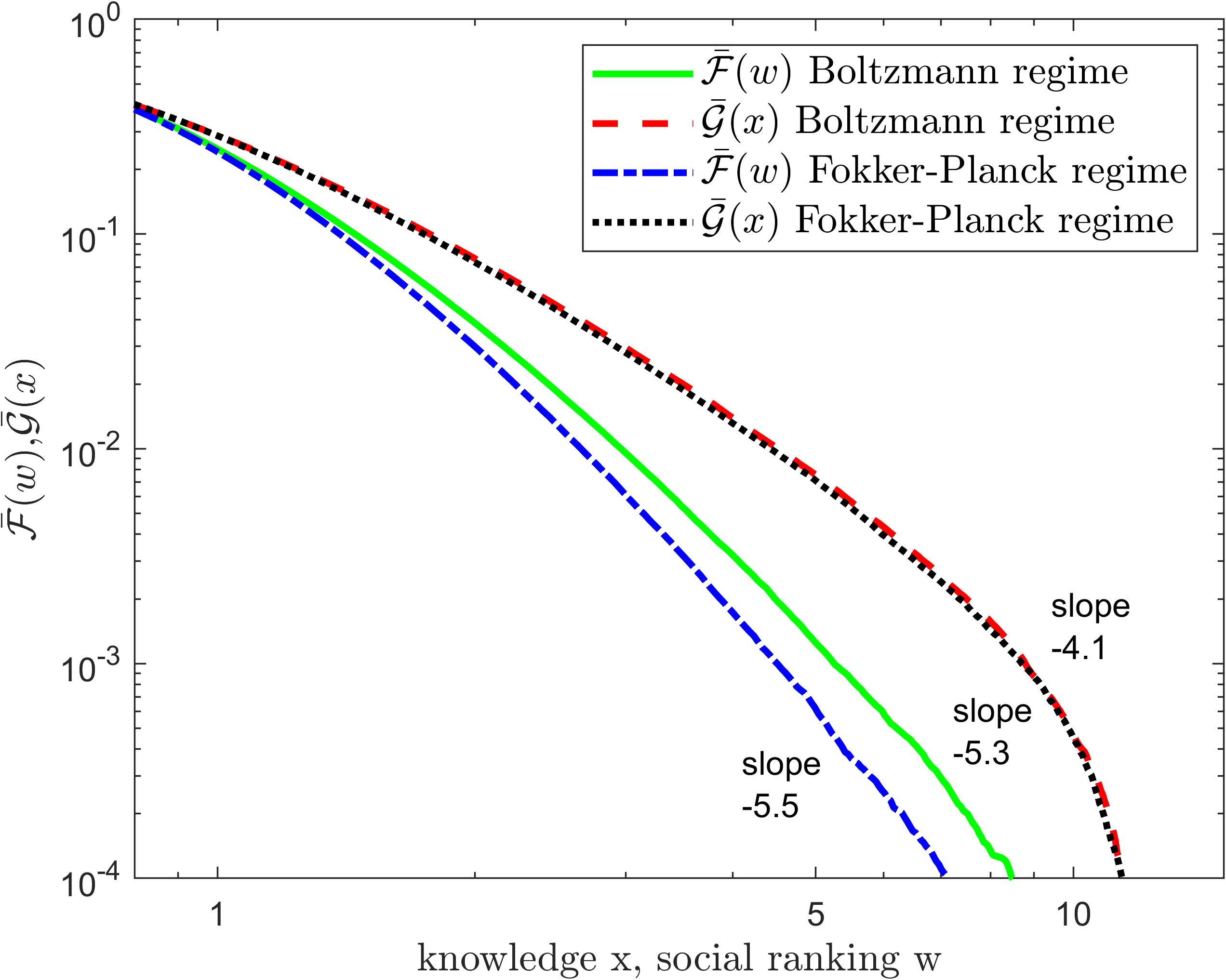}  \\  
        %\end{tabular}
        \caption{Test 1. Left image shows the two marginal densities in the different regimes relative to knowledge and social status. Right image shows the complementary cumulative marginal distributions for the same quantities in log-log scale. The slopes $p$ of the tails are estimated from $\mathcal{\bar{F}}(w)\approx w^{-p}$ and $\mathcal{\bar{G}}(x)\approx x^{-p}$ for larger $w$ and $x$.
        }
        \label{fig:test2}
    \end{center}
\end{figure}

In Figure \ref{fig:test1}, we show the results for the kinetic density in the above defined setting. In the same figure, we also report the distribution of a random subset of particles composed of $N=1000$ individuals at equilibrium showing a possible realization of the coupled model detailed in Section \ref{model-full}. These results are compared with the case in which the formation of the social structure and the knowledge/education are considered independent. In Figures \ref{fig:test2}, we plot instead the marginal densities together with the tail distributions as defined in equation \eqref{disg}-\eqref{disf} and \eqref{cumg}-\eqref{cumf}. The complementary cumulative distribution are shown in log--log scale on the right to better visualize the tails behaviour and determine the slopes of the polynomial decay which, as foreseen, are observed. 

The same test is subsequently performed in a Fokker--Planck regime, scaling all interaction parameters by a factor $10$. This corresponds to a scaling factor $\epsilon = 0.01$. The final computation time is the same, but the time step is now chosen as $\Delta t = 0.01$. This choice of the scaling parameter is enough to observe the convergence to the theoretical steady state distribution recalled in equations \eqref{equilibrio}-\eqref{equilibrio2} when the two dynamics are fully uncoupled. The results are shown again in Figures \ref{fig:test2}-\ref{fig:test3} and \ref{fig:test4}. They indicate a good agreement with the Boltzmann description. The major differences between the two scaling are noted in the peak of the distribution and in the low knowledge and social status regions. However, the two distributions remain very close each other for the two regimes. 

From the depicted results, it is clear that the effect of knowledge in minimizing the interval of convexity of the value function and in diminishing the variance of the stochastic interactions for highly educated individuals produces a tendency for people with an average or high level of knowledge to occupy the middle class region. Moreover, one can state from this numerical experiment that it is very difficult to reach a high level of knowledge for individuals belonging to the working class.

\subsection{Test 2}
\begin{figure}[!htbp]
    \begin{center}
%       \begin{tabular}{cc}    \hspace{-0.5cm}
            \includegraphics[width=0.45\textwidth]{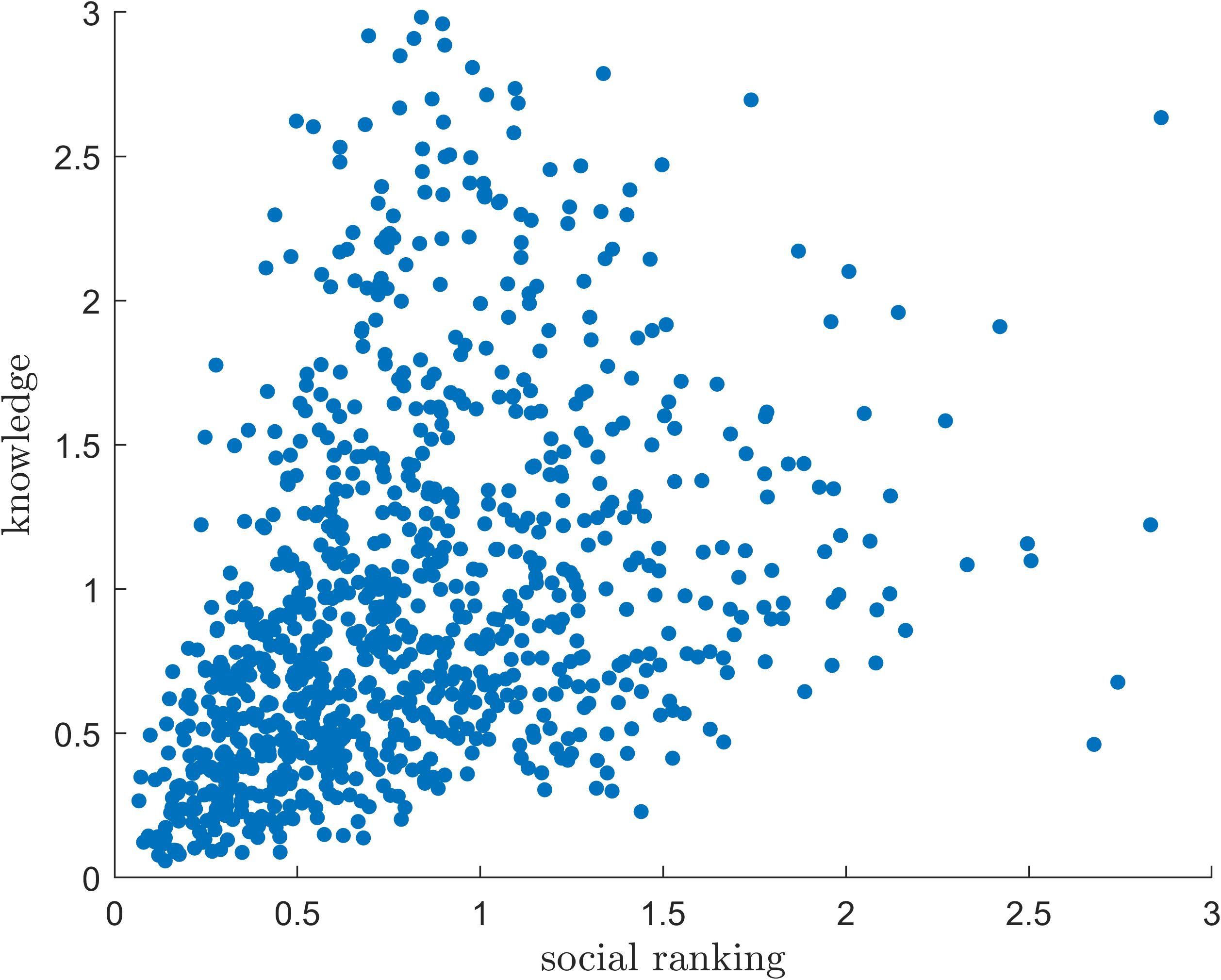} 
           \hspace{0.7cm}
            \includegraphics[width=0.45\textwidth]{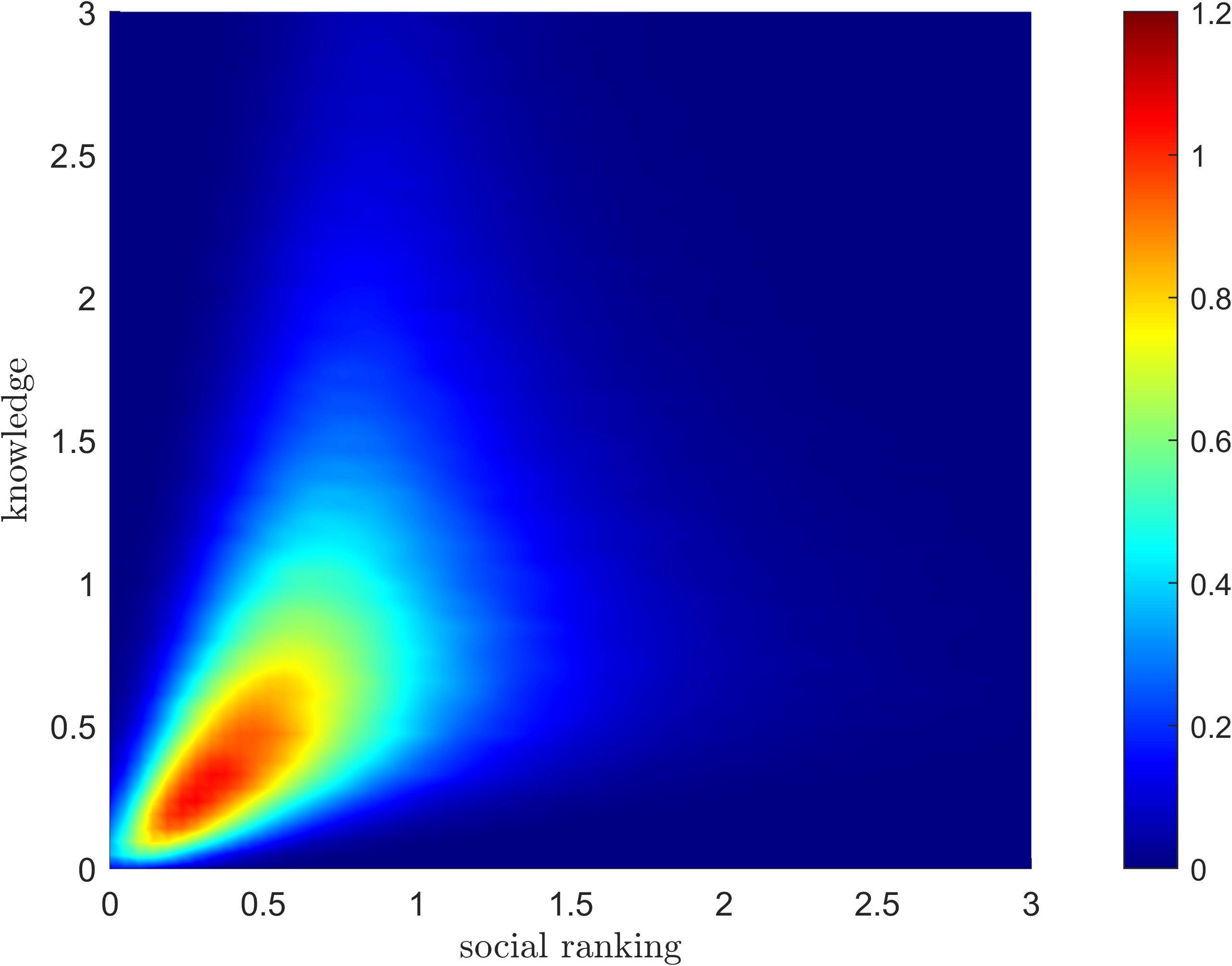}   \\  
%       \end{tabular}
        \caption{Test 2: particles solution with N = 1000 agents (left) and the kinetic density (right). The social structure depends upon knowledge as defined in \eqref{alfa} and \eqref{beta} while the knowledge depends upon the position occupied in the social hierarchy through \eqref{kn1}.}
        \label{fig:test3}
    \end{center}
\end{figure}

\begin{figure}[!htbp]
    \begin{center}
    %   \begin{tabular}{cc}    \hspace{-0.5cm}
            \includegraphics[width=0.45\textwidth]{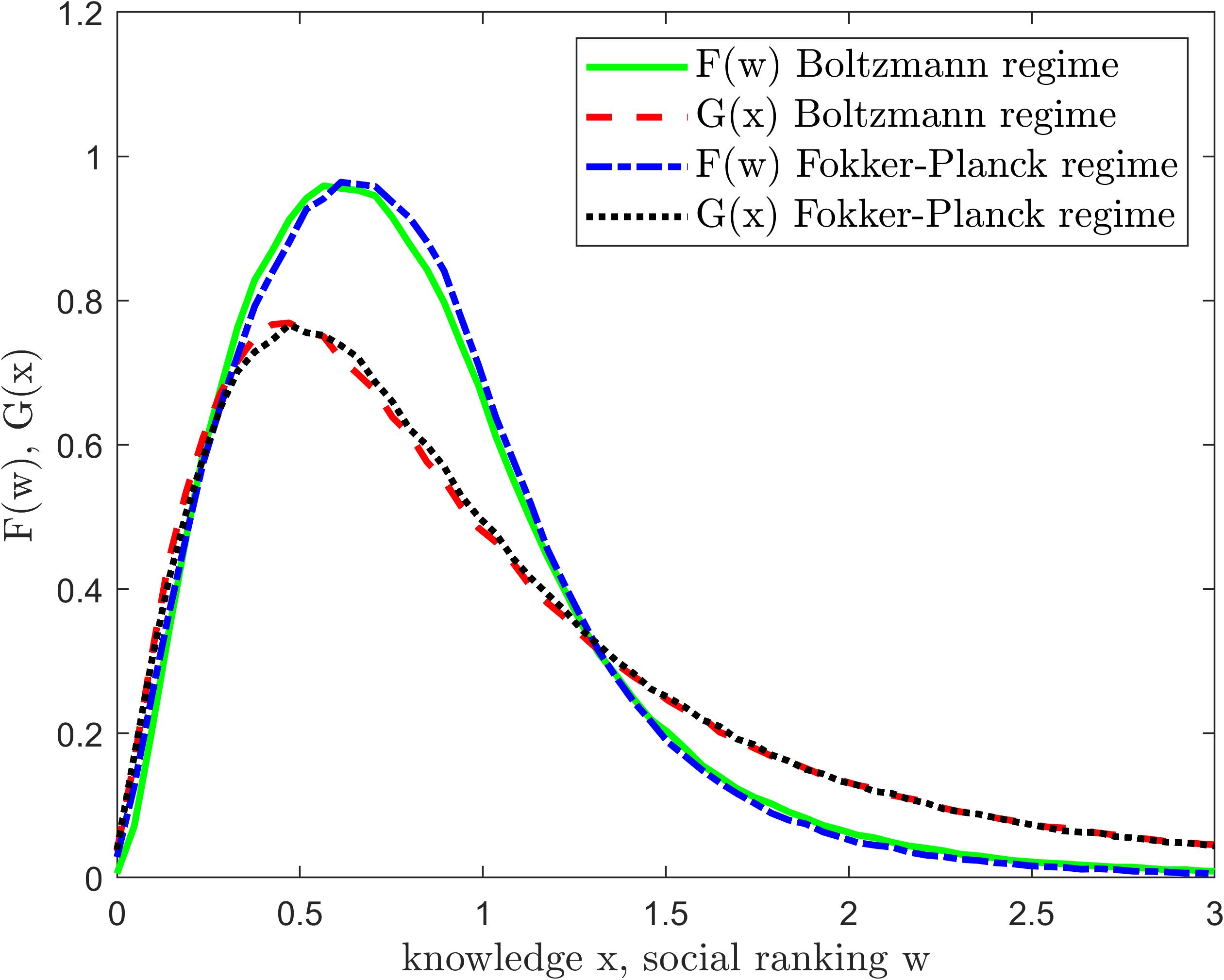} 
            \hspace{0.7cm}
            \includegraphics[width=0.45\textwidth]{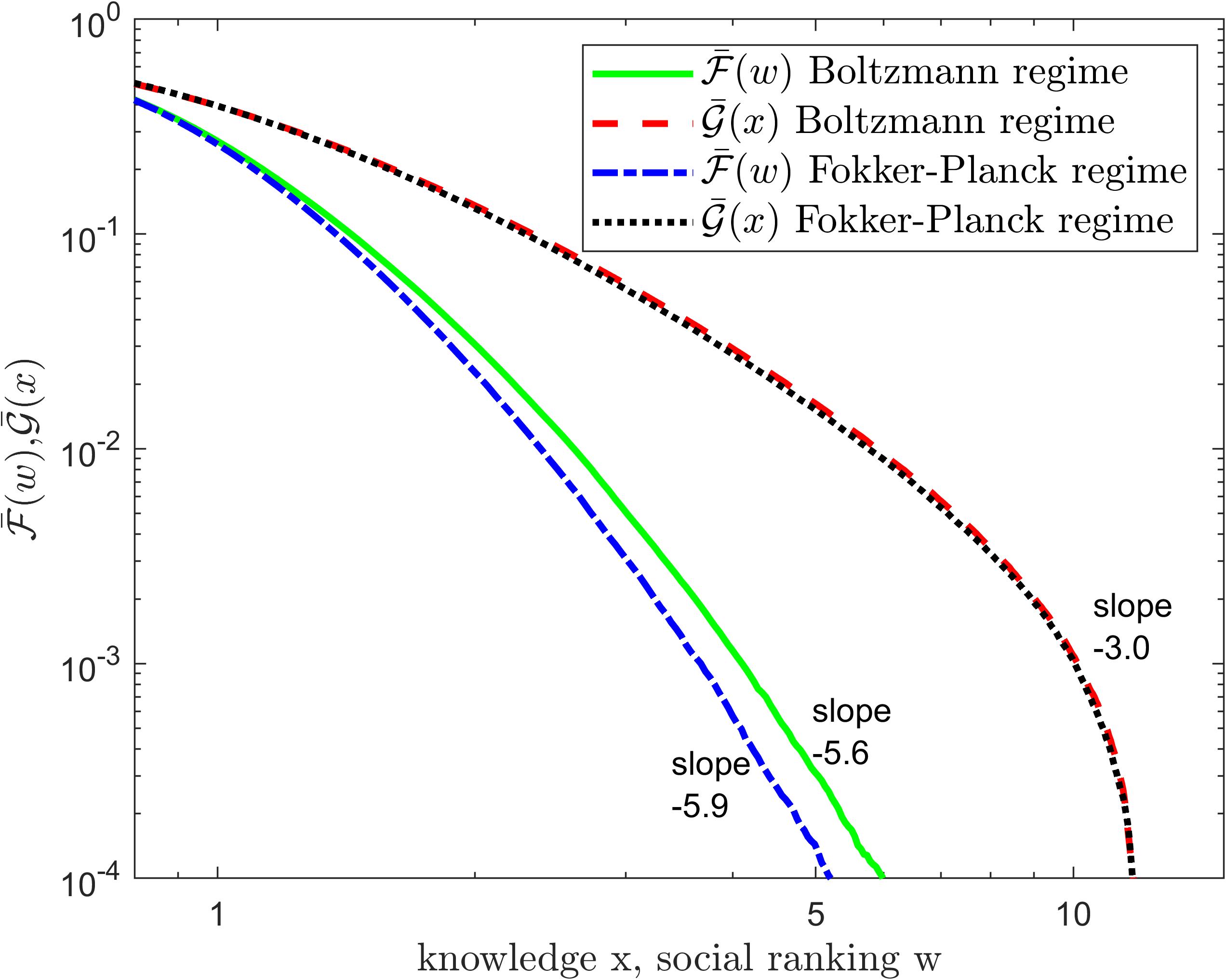}   \\  
    %   \end{tabular}
        \caption{Test 2. Left image shows the two marginal densities relative to knowledge and social status in the different regimes. Right image shows the complementary cumulative marginal distributions for the same quantities in log-log scale. The slopes $p$ of the tails are estimated from $\mathcal{\bar{F}}(w)\approx w^{-p}$ and $\mathcal{\bar{G}}(x)\approx x^{-p}$ for larger $w$ and $x$.}
        \label{fig:test4}
    \end{center}
\end{figure}

In this second test, we take into account that knowledge may depend upon the position occupied by individuals in the social ladder. Thus, the microscopic interaction used to construct the Boltzmann-type dynamics is modified for what concerns only the knowledge through equation \eqref{kn1} which replaces equation \eqref{kn} where $E(w)$ is defined in \eqref{edu}. The other parameters of the model are kept unchanged  More precisely, we assume
\be
E(w) = \frac{m+Mw^\xi}{1 + w^\xi}=\frac{0.1+3w^2}{1 + w^2},
\ee
while the same values of the Test 1 for $\mu$, $\bar w_L$ for the social climbing model and $\lambda,\lambda_B,\nu,M_B$ in the knowledge model are used. The results are reported in Figure \ref{fig:test3} for the full density and in Figures \ref{fig:test4} for the marginal densities and their tail distributions. As before, two different regimes are considered, namely the Boltzmann with $\epsilon=0.1$ and the Fokker-Planck regime with $\epsilon=0.01$, the two giving similar results as shown in the images. Comparing the results obtained with the ones of the first test, here we can observe that there is a larger shift of the population towards higher positions in the social ladder but especially for individuals possessing a larger knowledge. In particular, the tail of the marginal density for what concerns the knowledge is fatter if compared with the first test case.

%\begin{figure}[!htbp]
%   \begin{center}
%       \begin{tabular}{cc}    \hspace{-0.5cm}
%           \includegraphics[width=0.5\textwidth]{figures/test2/test1_pbis} 
%           \hspace{0.5cm}
%           \includegraphics[width=0.5\textwidth]{figures/test2/test1_bis}   \\  
%       \end{tabular}
%       \caption{Test 2: particles solution with N = 1000 agents (left) and the kinetic density (right). The social structure depends upon knowledge as defined in \eqref{alfa} and \eqref{beta} while the knowledge depends upon the position occupied in the social hierarchy through \eqref{kn1}.}
%       \label{fig:test3}
%   \end{center}
%\end{figure}
%
%\begin{figure}[!htbp]
%   \begin{center}
%       \begin{tabular}{cc}    \hspace{-0.5cm}
%           \includegraphics[width=0.5\textwidth]{figures/test2/test1_marginals_3} 
%           \hspace{0.5cm}
%           \includegraphics[width=0.5\textwidth]{figures/test2/test1_marginals_tails_3}   \\  
%       \end{tabular}
%       \caption{Test 2. Left image shows the two marginal densities relative to knowledge and social status. Right image shows the complementary cumulative marginal distributions for the same quantities in log-log scale. The slopes $p$ of the tails are estimated from $\mathcal{\bar{F}}(w)\approx w^{-p}$ and $\mathcal{\bar{G}}(x)\approx x^{-p}$ for larger $w$ and $x$.}
%       \label{fig:test4}
%   \end{center}
%\end{figure}

%\begin{figure}[!htbp]
%   \begin{center}
%       \begin{tabular}{cc}    \hspace{-0.5cm}
%           \includegraphics[width=0.5\textwidth]{figures/test1/mean_values_1_bis} 
%           \hspace{0.5cm}
%           \includegraphics[width=0.5\textwidth]{figures/test2/mean_values_1_bis}   \\  
%       \end{tabular}
%       \caption{The behaviour of the mean knowledge against the social status and of the mean social rank against knowledge. Test 1 left image, Test 2 right image. The Boltzmann and the Fokker-Planck scaling are shown for both cases.}
%       \label{fig:test5}
%   \end{center}
%\end{figure}

\begin{figure}[!htbp]
    \begin{center}
    %   \begin{tabular}{cc}    \hspace{-0.5cm}
            \includegraphics[width=0.45\textwidth]{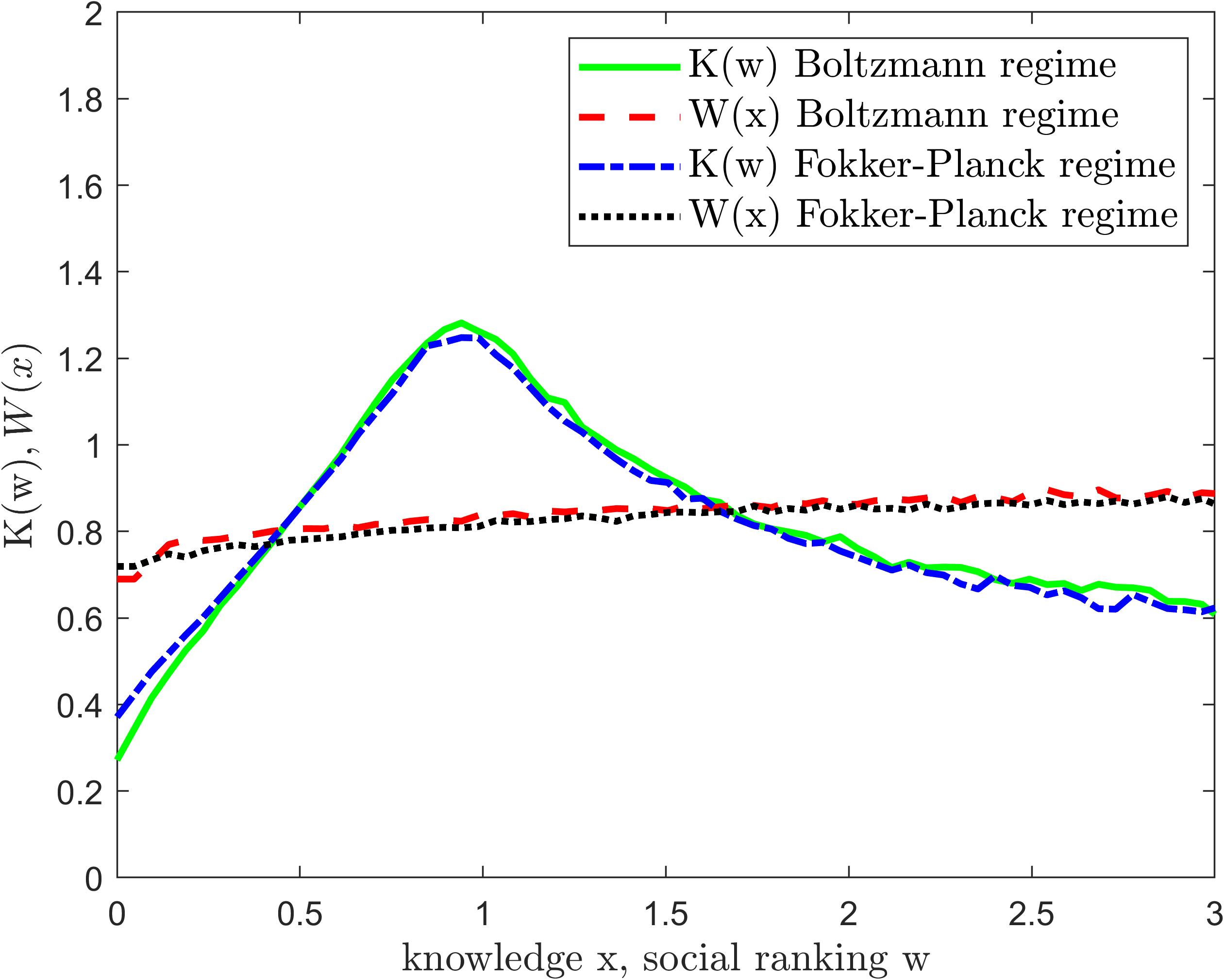} 
            \hspace{0.7cm}
            \includegraphics[width=0.45\textwidth]{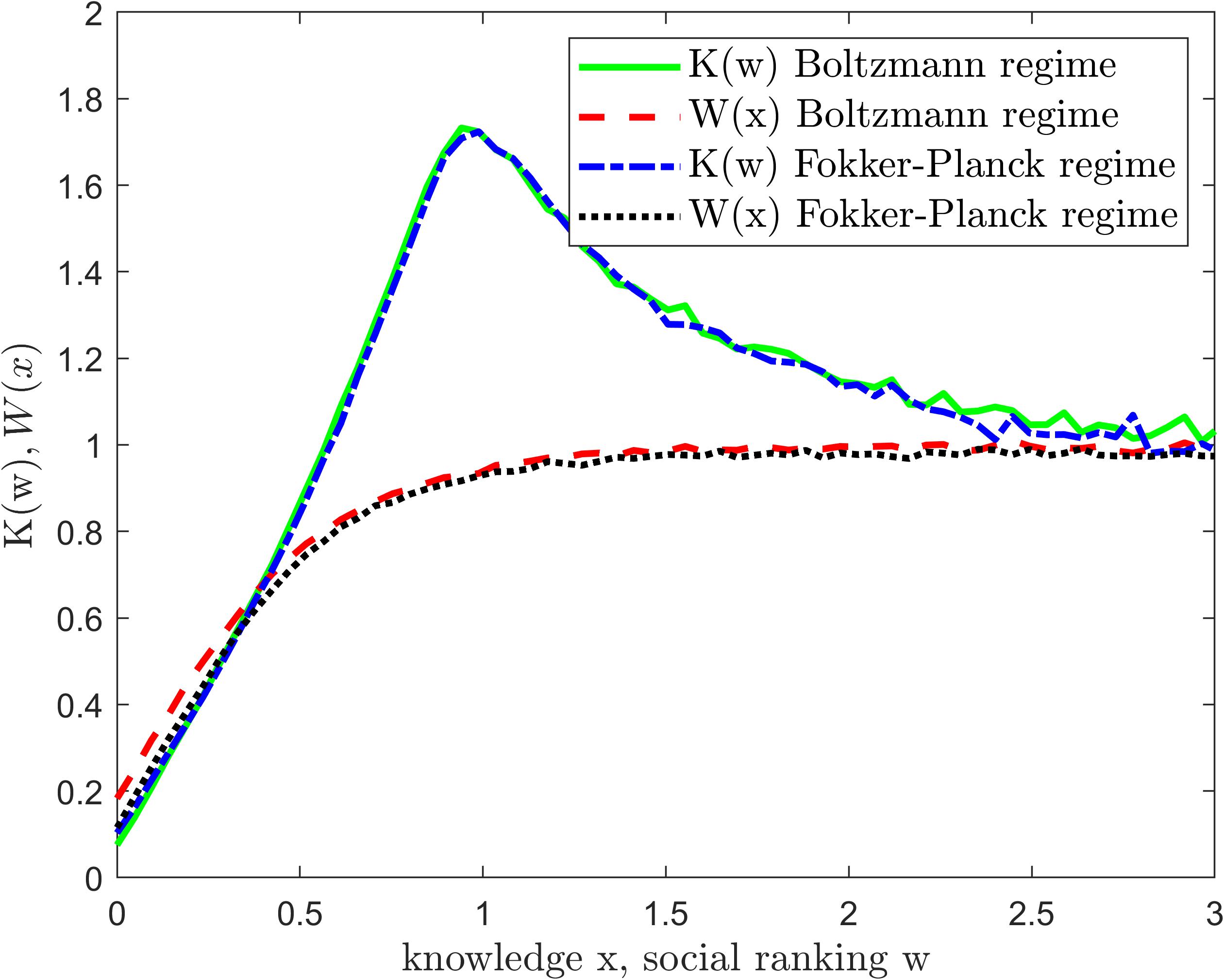}   \\  
    %   \end{tabular}
        \caption{The behaviour of the mean knowledge against the social status and of the mean social rank against knowledge. Test 1 left image, Test 2 right image. The Boltzmann and the Fokker-Planck scaling are shown for both cases.}
        \label{fig:test5}
    \end{center}
\end{figure}

As a second measure of the phenomenon under consideration, we report the profiles of the local mean value of the social status and of the local mean knowledge. These are defined as
\be
\begin{split}
W(x,t)&= \frac{1}{G(x,t)}\int f(x, w, t)w dw, \\
 K(w, t) &= \frac{1}{F(w, t)}\int f(x, w, t)x dx.
\end{split}
\ee
The profiles are shown in Figure \ref{fig:test5} for the Boltzmann and the Fokker-Planck regime when the steady state is reached. They show that for the first test, the mean social status is essentially independent with respect of the knowledge except for individuals having a very low level of education. This effect is amplified for the second test case where the education has a larger impact on the social status. In this second case the larger is the education level the larger is the average social status. However, a saturation of this index is observed for large values of the knowledge. The second main outcome is a pick of the average educational level for agents belonging to the middle classes. This effect is amplified when the knowledge is additionally a function of the social status (Test 2). 

\subsection{Test 3}
\begin{figure}[!htbp]
    \begin{center}
    %   \begin{tabular}{cc}    \hspace{-0.5cm}
            \includegraphics[width=0.45\textwidth]{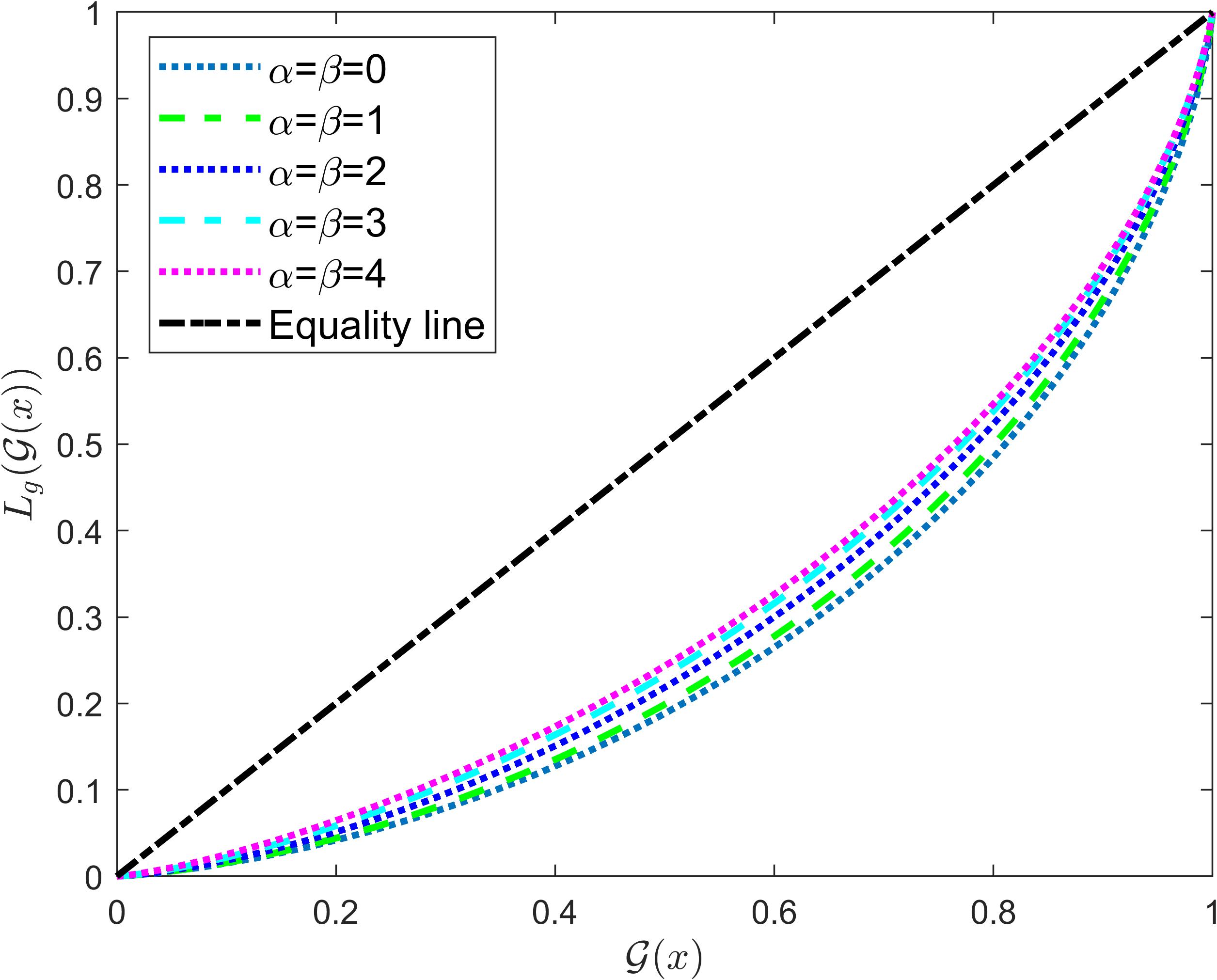}
            \hspace{0.7cm}
            \includegraphics[width=0.45\textwidth]{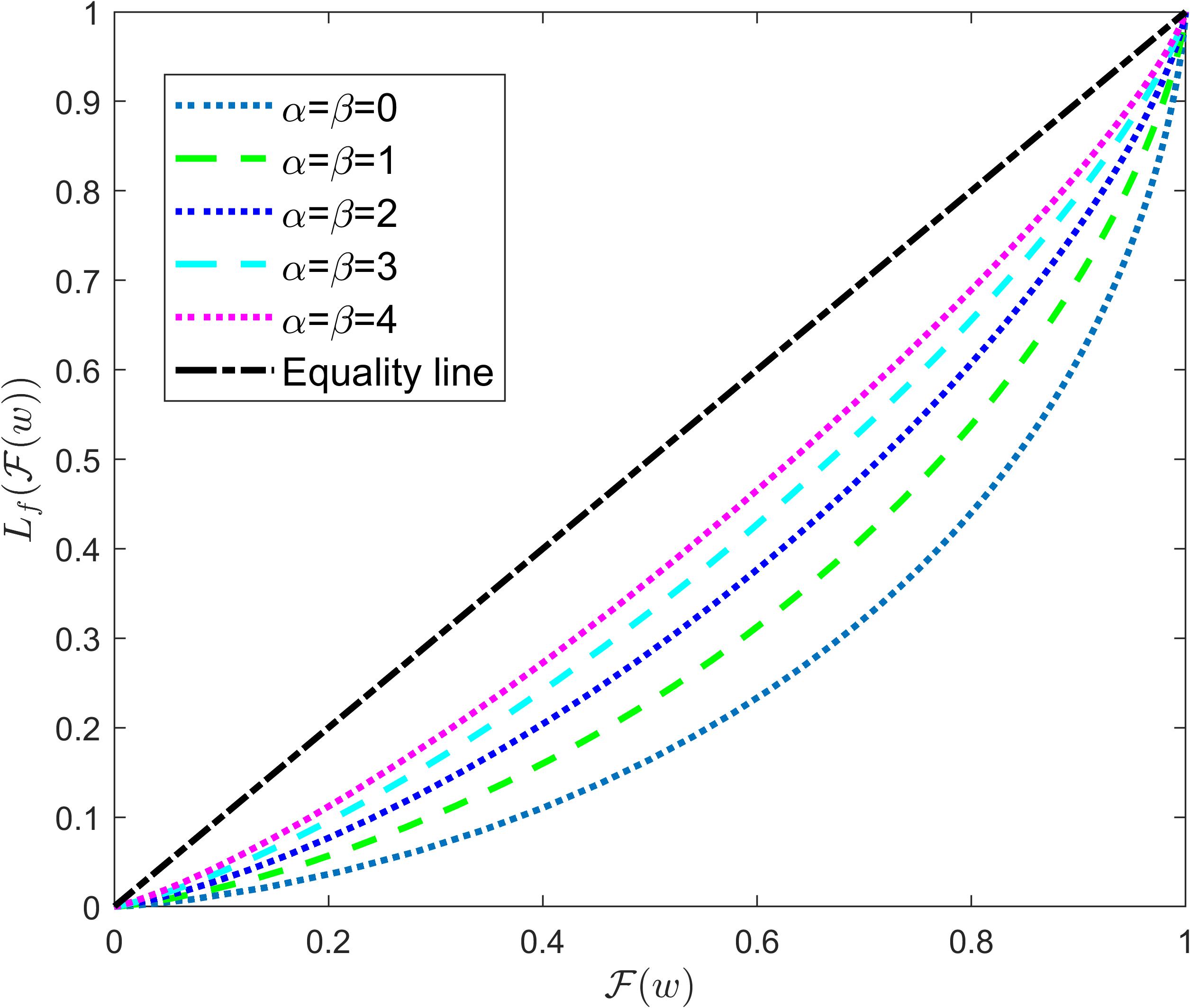}  \\  
                \includegraphics[width=0.45\textwidth]{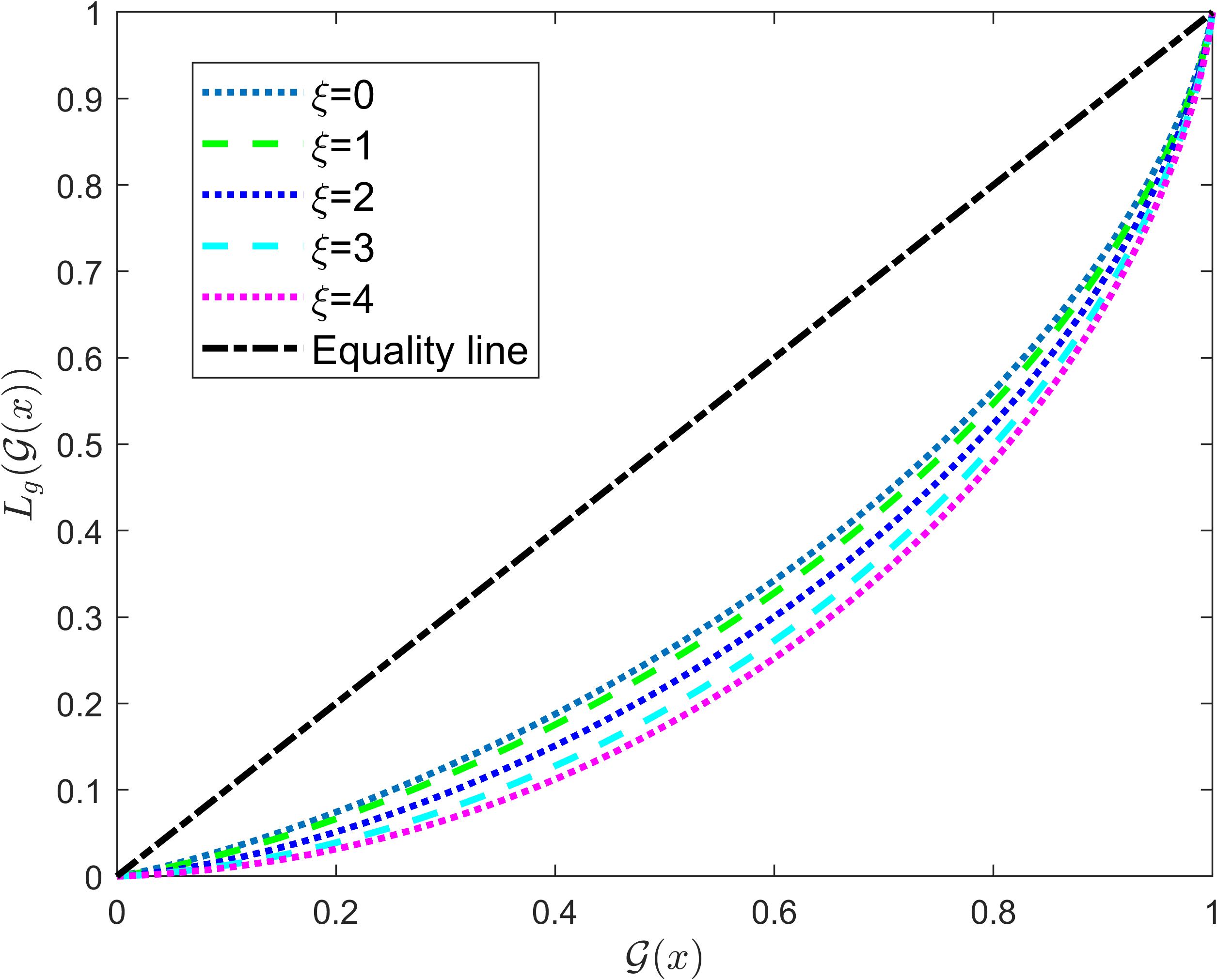}
            \hspace{0.7cm}
            \includegraphics[width=0.45\textwidth]{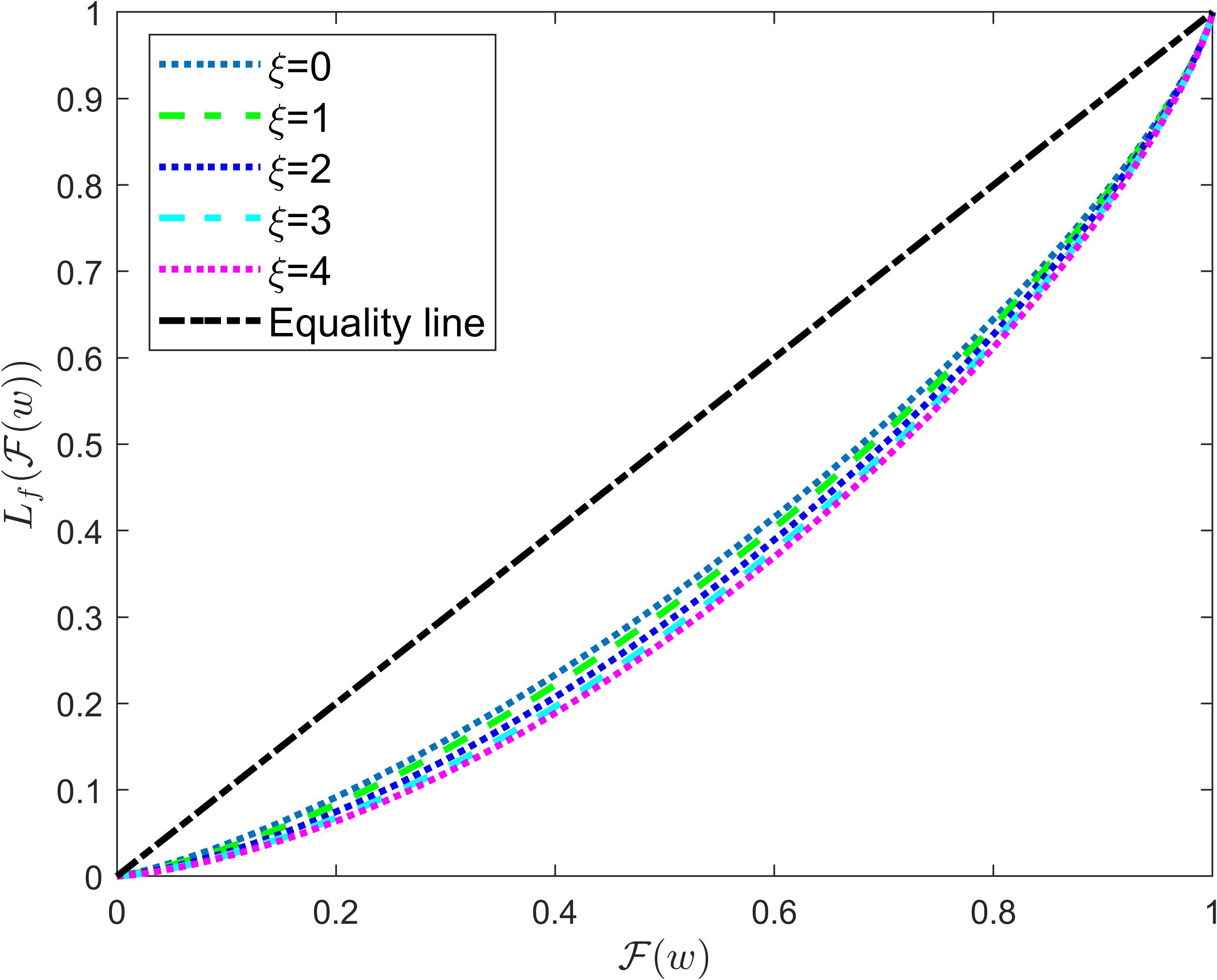} 
%       \end{tabular}
        \caption{Top: the behavior of the Lorenz curves for the social stratification and the knowledge for different values of $\alpha,\beta$ and for fixed $\xi$ in equations \eqref{alfa},\eqref{beta},\eqref{edu}. For the left image the Gini coefficients span from $\Gamma_x = 0.48$ for $\alpha=\beta=0$ to $\Gamma_x = 0.38$ for $\alpha=\beta=4$. For the right image the Gini coefficients span from $\Gamma_w = 0.54$ for $\alpha=\beta=0$ to $\Gamma_w = 0.16$ for $\alpha=\beta=4$.
        Bottom: the behavior of the Lorenz curves for the social stratification and the knowledge for different values of $\xi$ and fixed $\alpha,\beta$ in equations \eqref{alfa},\eqref{beta},\eqref{edu}. For the left image the Gini coefficients span from $\Gamma_x = 0.26$ for $\xi=0$ to $\Gamma_x = 0.32$ for $\xi=4$. For the right image the Gini coefficients span from $\Gamma_w = 0.35$ for $\xi=0$ to $\Gamma_w = 0.48$ for $\xi=4$.
        }
        \label{fig:test6}
    \end{center}
\end{figure}

In order to gain a deeper understanding of the phenomenon and to have a more detailed view of the emerging steady states, we finally show the Lorenz curve and the Gini index computation in this last part (cf.  \cite{CCCC,DPT} for details about this index). When the Gini index is used to study the wealth distribution, its value should be understood as a measure of a country's inequality. Here, we perform a similar analysis but in terms of the education level and of the position occupied in the social hierarchy by individuals, i.e. the possible emergence of social elites. By definition, the values taken by the Gini index varies in $[0, 1]$ where the value $0$ indicates that the society is equal while the value $1$ expresses perfect inequality. In this latter case, we would observe a fat tail in the distribution of the social status or of the knowledge distribution indicating the presence of an elite which causes the Gini index to grow. In the case of the wealth measure in modern economies, one often observes a value between $[0.2, 0.5]$ for this indicator \cite{DY01}. This index is computed starting from the Lorenz curves defined as
\be\label{GINI}
L_f(\mathcal{F}(w))=\frac{\int_0^w F^\infty(w_*)w_*dw_*}{\int_0^\infty F^\infty(w_*)w_*dw_*}
\ee
and
\be\label{GINI1}
L_g(\mathcal{G}(x))=\frac{\int_0^w G^\infty(x_*)x_*dx_*}{\int_0^\infty G^\infty(x_*)x_*dx_*}
\ee
where the marginal distributions $F^\infty(w)$ and $G^\infty(x)$ correspond to the steady state solutions. From the above defined quantities, it is possible to recover the Gini coefficients as
\be
\Gamma_w=1-2\int_0^1 L_f(\mathcal{F}(w)) dw, \quad \Gamma_x=1-2\int_0^1 L_g(\mathcal{G}(x)) dx.
\ee
In Figure \ref{fig:test6} we show the Lorenz curves obtained in the same setting of Test 1 and Test 2 and in the Boltzmann regime. In particular, the figure shows the Lorenz curves corresponding to \eqref{GINI1} on the top left and to \eqref{GINI} on the top right for different values of the exponents $\alpha$ and $\beta$ in respectively equation \eqref{alfa} and \eqref{beta}. These are used respectively to decrease the convexity interval in the value function \eqref{vff} and to modify the stochasticity in the formation of the social structure while $\xi$ is fixed to $\xi=2$. We recall in particular that $\sigma(x)$ has been designed in order to reduce the randomness present in the social climbing model when the education level becomes particularly high while larger values of $\beta$ enhance this behavior. The results show that when one considers a model in which the social status is influenced by education and vice-versa a reduction of the inequality can be reached in terms of social stratification. Indeed, in the uncoupled case, one can observe that the tails become thinner when the variance decreases indicating that a larger level of equality may be reached by the system. The Gini coefficient jumps from a value of $0.52$, corresponding to the fully decoupled case, to a value $0.16$ for the social stratification. On the other hand, this coupling effect has a smaller impact in modifying the inequalities in the knowledge/education distribution. In fact, in the case of the $L_g(\mathcal{G}(x))$, the Lorenz curves corresponding to knowledge are close each other, still it is possible to appreciate a difference in terms of the Gini coefficient which passes from $\Gamma_x = 0.47$ to $\Gamma_x = 0.38$ for larger values of $\beta$ and $\alpha$. 

The images on the bottom of Figure \ref{fig:test6} show the Lorenz curves when $\alpha=\beta=2$ and $\xi$ varies in $[0,4]$. From these images, one can see that increasing $\xi$ implies larger values of the Gini index both for knowledge as well as for social status, even if this effect is less apparent for this latter. Thus one can infer that as $\xi$ grows and thus the education level largely depends on the status of individuals then inequalities increases as opposite to the previous case. This result can be interpreted saying that if education depends upon social status, people belonging to the social elites has more possibilities to get better education and thus larger possibilities to climb position in the social ladder thus increasing inequalities.

\begin{figure}[!htbp]
    \begin{center}
        %   \begin{tabular}{cc}    \hspace{-0.5cm}
        \includegraphics[width=0.45\textwidth]{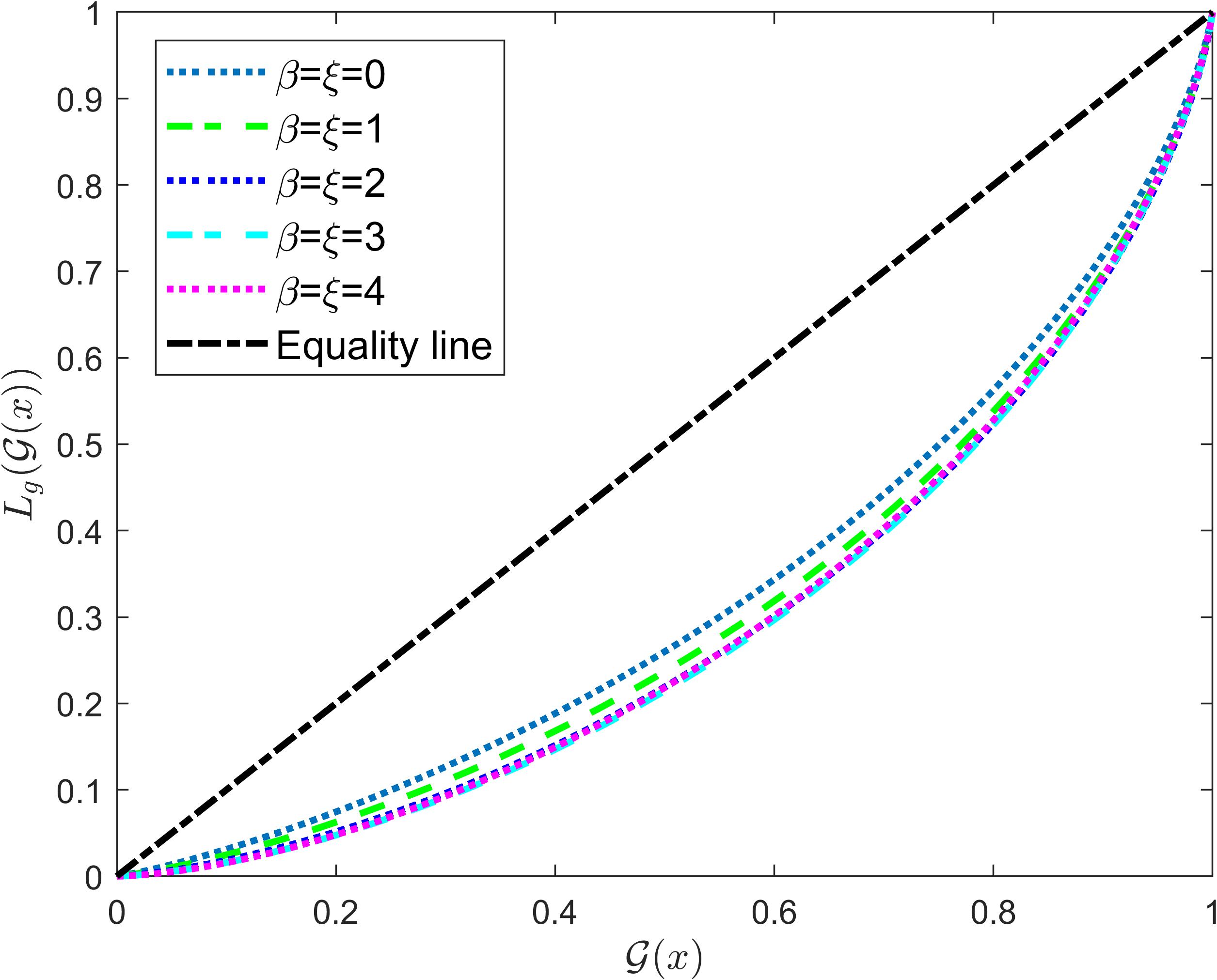}
                    \hspace{0.7cm}
        \includegraphics[width=0.45\textwidth]{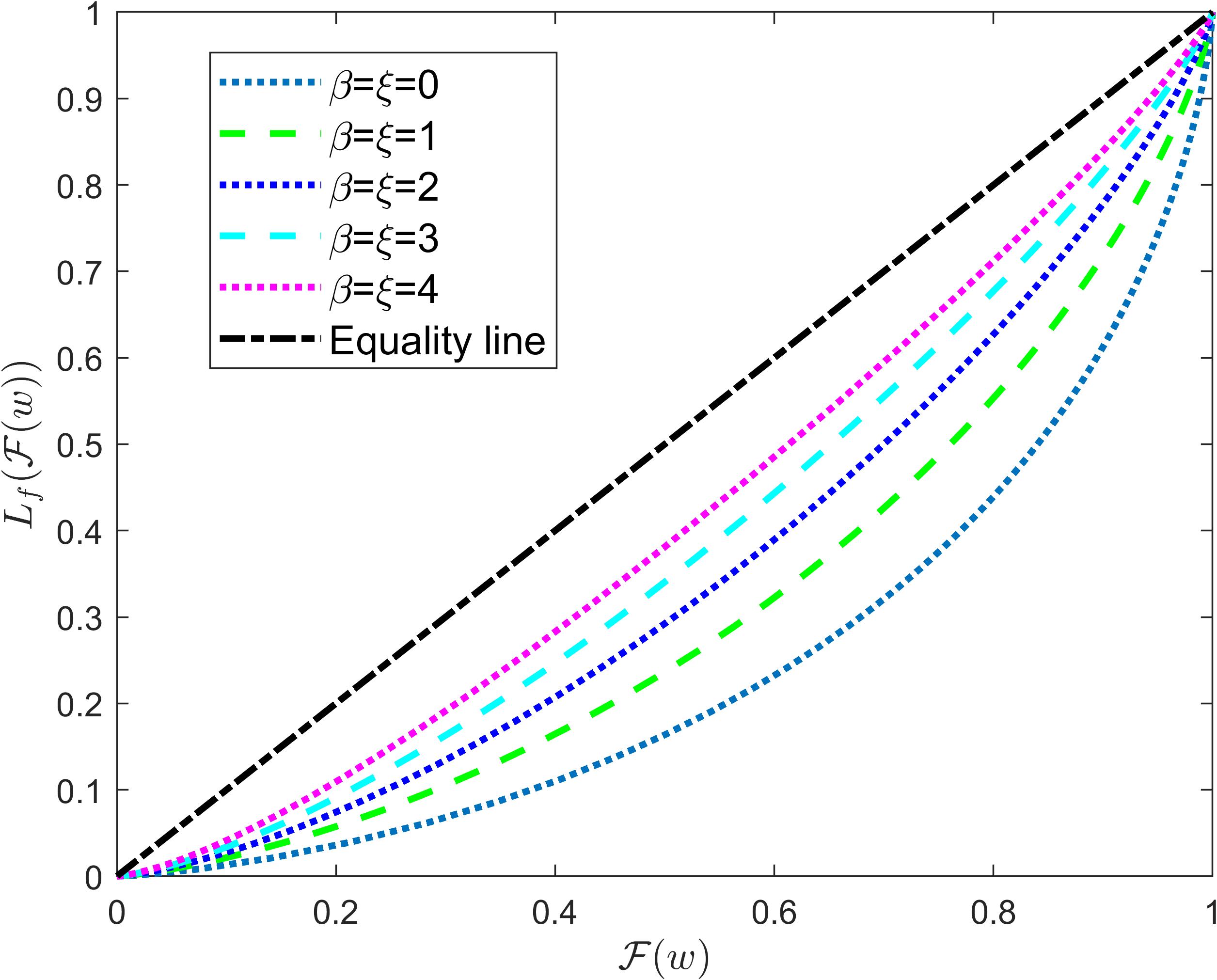}  \\  
        \includegraphics[width=0.45\textwidth]{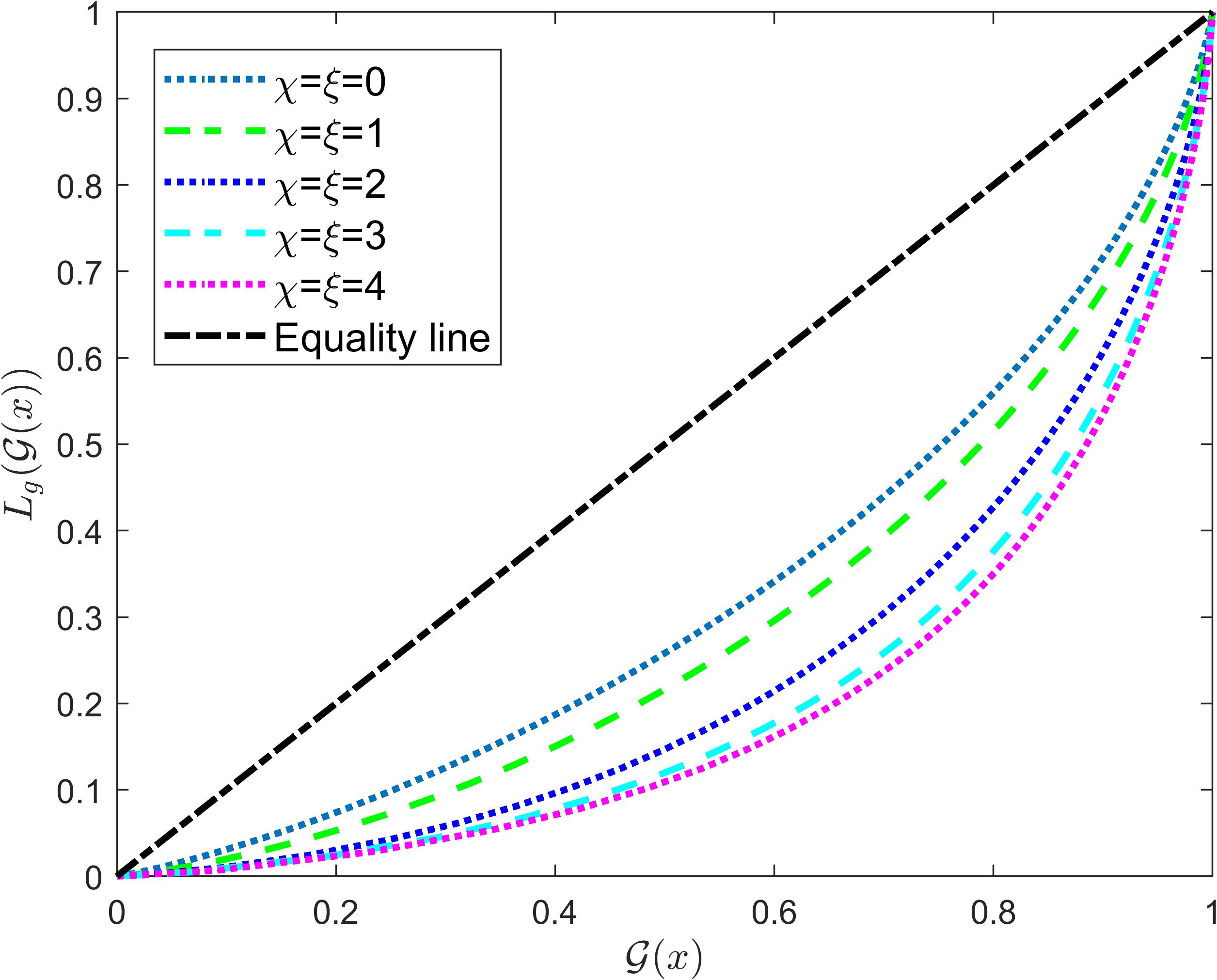}
                    \hspace{0.7cm}
        \includegraphics[width=0.45\textwidth]{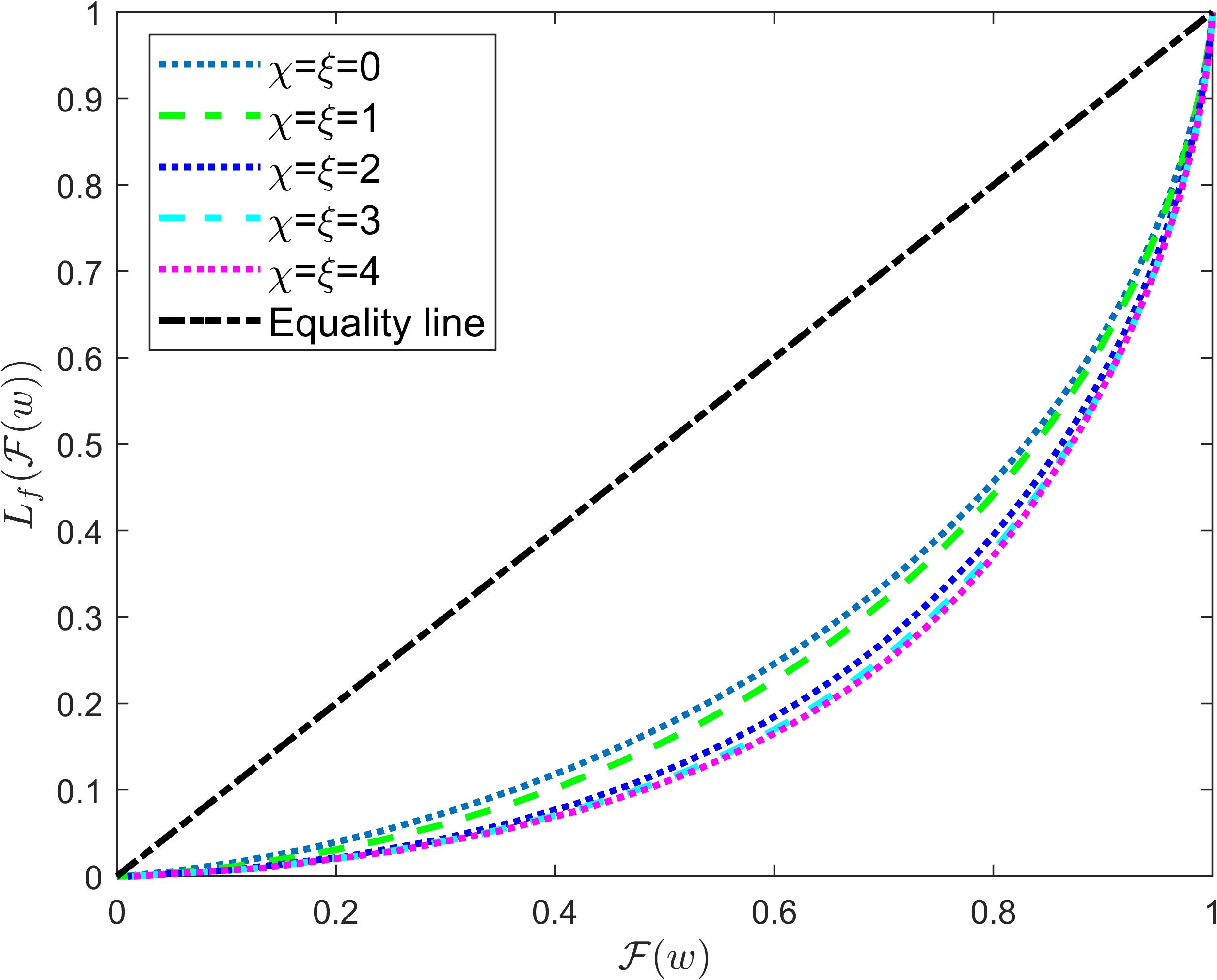} 
        \includegraphics[width=0.45\textwidth]{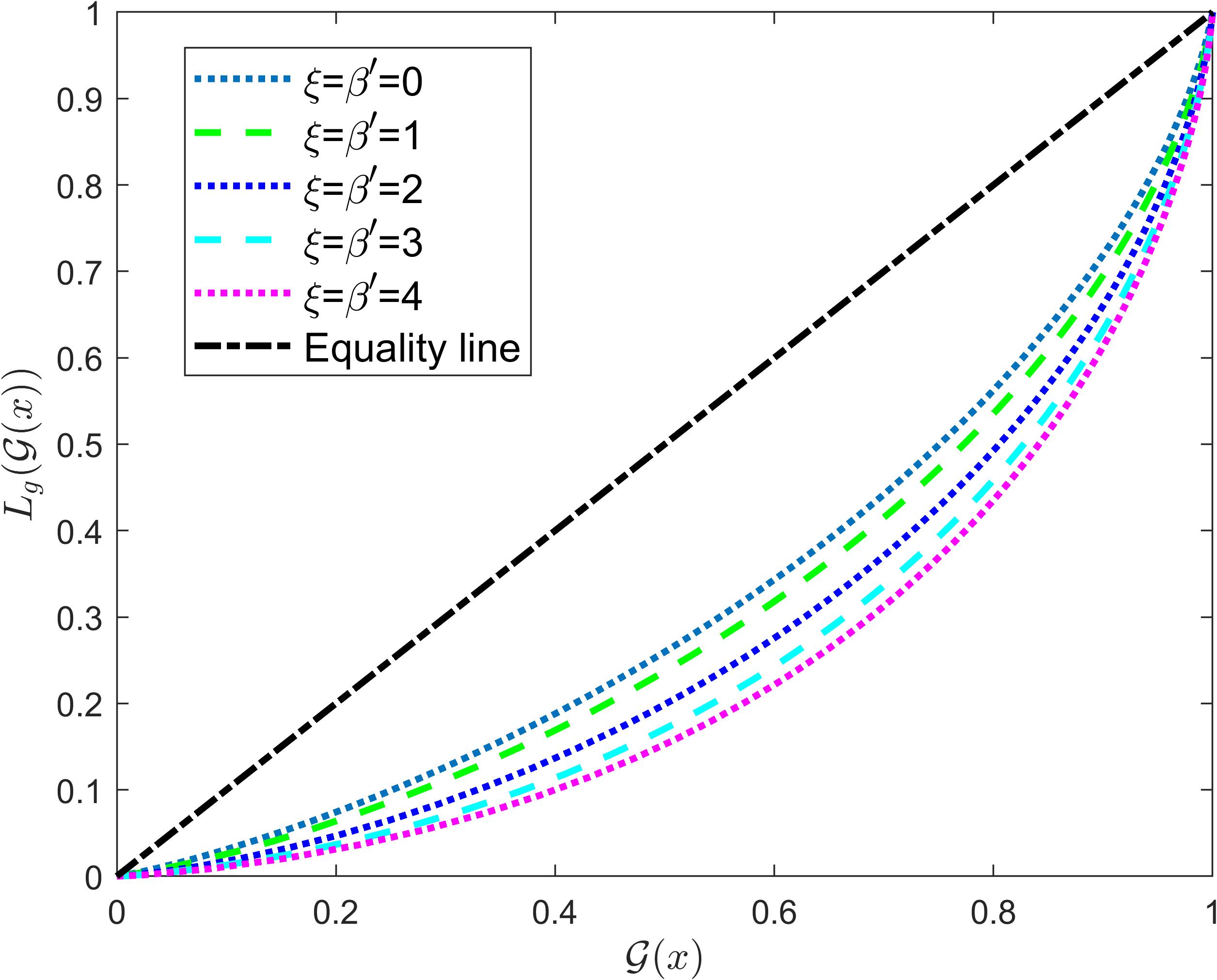}
        \hspace{0.7cm}
        \includegraphics[width=0.45\textwidth]{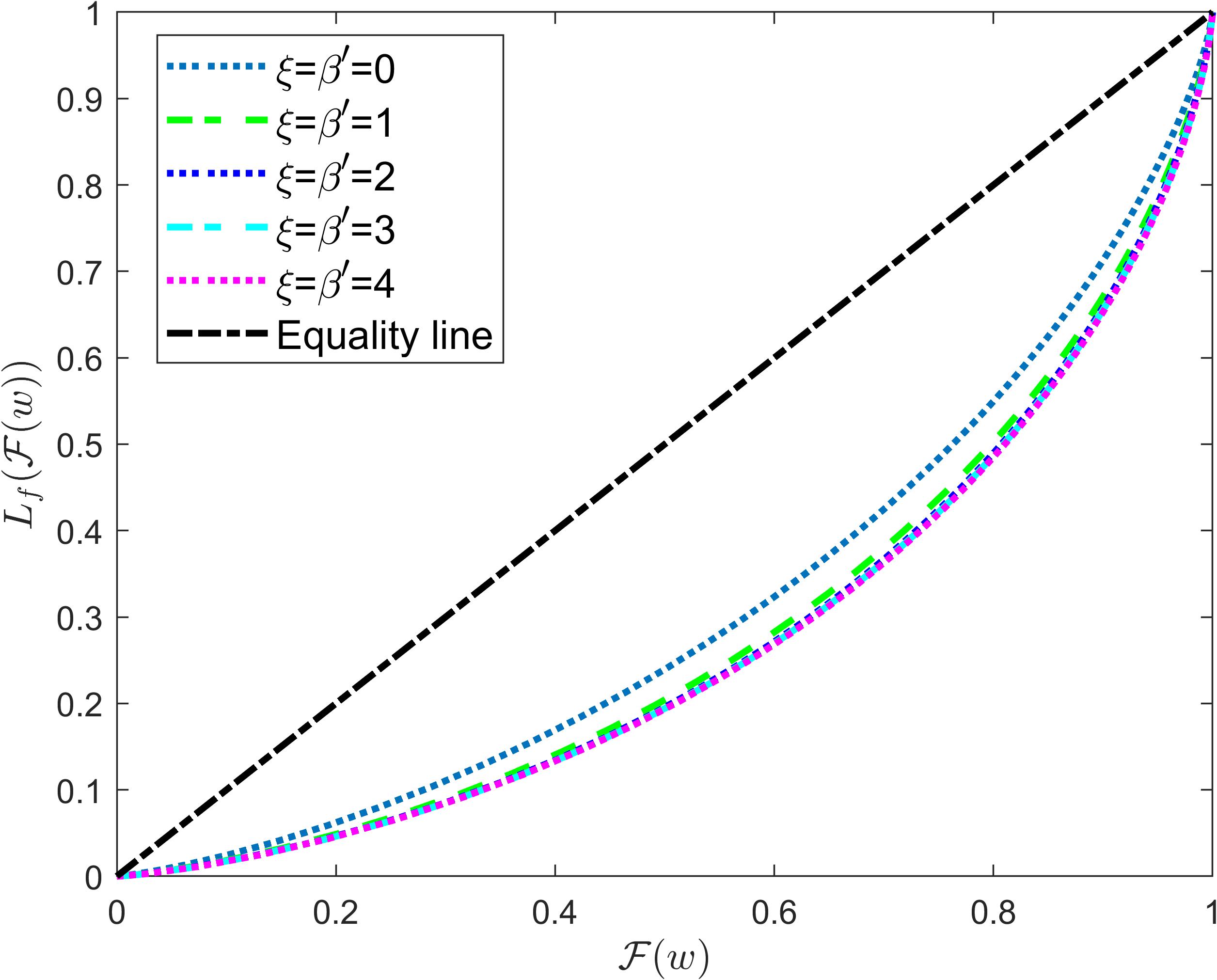} 
        %       \end{tabular}
        \caption{The behavior of the Lorenz curves for the social stratification and the knowledge. Top: as a function of $\beta$ and $\xi$ as defined in \eqref{beta} and \eqref{edu} with fixed $\alpha$ as defined in \eqref{alfa}. Middle: as a function of $\chi$ and $\xi$ as defined in \eqref{beta1} and \eqref{edu} and fixed $\alpha$ as defined in \eqref{alfa}. 
            Bottom: as a function of $\beta^\prime$ and $\xi$ as defined in \eqref{beta2} and \eqref{edu} and fixed $\alpha$ as defined in \eqref{alfa}. 
        }
        \label{fig:test7}
    \end{center}
\end{figure}

Finally, in Figure \ref{fig:test7} we show on the top the Lorenz curves for the case in which $\alpha$ is fixed and $\beta$ and $\xi$ vary, the rest of parameters remaining unchanged. In this case, we observe as for the case of $\xi$ fixed and $\alpha,\beta\in [0,4]$ a reduction of the inequalities for the social stratification while knowledge sees inequality to grow as the effect of this coupling grows even if education looks much less influenced in this situation by the coupling conditions. This setting combines the positive effects of the reduction of the variance when individuals increase their education level with the negative, even if realistic, effects of relating the possibilities to be educated with their means. 

The last two rows in Figure \ref{fig:test7} explore a different setting. The middle row reports the case in which \eqref{beta1} is used instead of \eqref{beta2} for coupling the stochastic behavior in the formation of the social hierarchy with the knowledge level, the other quantities remaining unchanged. In details the function used is
\be
\sigma(x) = \frac{0.01+0.9x^{-\chi}}{1 + x^{-\chi}}
\ee   
meaning that the variance may assume values between $0.01$ and $0.9$ and that the effect of this coupling increases as $\chi=[0,4]$ increases. We notice that for this case the behaviors of the curves are qualitatively analogous of the ones obtained with $\sigma(x)$ fixed by \eqref{beta} and both $\beta$ and $\xi$ varying in the interval $[0,4]$ (see bottom of Figure \ref{fig:test6}) even if quantitatively the inequalities are much pronounced in this situation as the coupling between the two phenomena becomes stronger. 

To conclude, we explore the case in which the variance in the social climbing model is defined in such a way that people with an average education level are the ones who are more subject to the unpredictability phenomena in the society being for uneducated and highly educated people the social mobility less probable. In this case we assume
\be
\sigma(x) = \frac {0.5}{1+(x-1)^{\beta^\prime}}
\ee
and we show the results for different values of the constant $\beta^\prime$ at the bottom of Figure \ref{fig:test7}, the other parameter being fixed as before. In this case, one can observe that education has a negative impact on the formation of the social hierarchy meaning that it contributes to enhance the difference in the society as for the case in which the variance has been assumed to vary as detailed in \eqref{beta1}, the same happens to the distribution of knowledge and its Lorenz curve.

%\begin{figure}[!htbp]
%    \begin{center}
%        \begin{tabular}{cc}    \hspace{-0.5cm}
%            \includegraphics[width=0.5\textwidth]{figures/test3/Lorentz2}
%            \hspace{0.5cm}
%            \includegraphics[width=0.5\textwidth]{figures/test3/Lorentz1}  \\  
%            \includegraphics[width=0.5\textwidth]{figures/test3/gini}
%        \end{tabular}
%        \caption{Top: the behavior of the Lorenz curves for the social stratification and the knowledge. For the left image the Gini coefficients span from $\Gamma_x = 0.48$ for $\alpha=\beta=0$ to $\Gamma_x = 0.38$ for $\alpha=\beta=4$. For the right image the Gini coefficients span from $\Gamma_w = 0.54$ for $\alpha=\beta=0$ to $\Gamma_w = 0.16$ for $\alpha=\beta=4$.
%            Bottom: the Gini coefficients for different values of $\beta$ and $\xi$ and fixed $\alpha=2$. 
%        }
%        \label{fig:test6}
%    \end{center}
%\end{figure}

\section{Conclusions}
We introduced and discussed a system of one-dimensional kinetic equations able to describe both the influence of higher education on social stratification of a multi-agent society, and the reverse influence of social stratification on the education of the collectivity. The system has been built by coupling a kinetic model for knowledge formation with a kinetic description of the social climbing. Two different cases have been analyzed in details. In the first case the elementary interaction characterizing the evolution of  the individual knowledge has been assumed independent of the individual social ranking. This assumption reproduces well the situation of countries where school education is overwhelmingly public, and the individual has the possibility of accessing almost any school, regardless of position on the social ladder. Second, we discussed the case in which the education level of an individual depends on the position occupied in the social ranking, a situation which more likely refers to countries where tuition fees are high at every level and 
% in which are present social inequalities in accessing higher education generally, 
become very high for more prestigious forms of higher education. 
Within this last assumption we obtained a fully coupled model in which knowledge and social status influence each other. Numerical experiments show that educational expansion is a socially progressive development, able to reduce socioeconomic inequalities. However, the more the education depends on the social status the more the inequalities in the society manifest themselves. Another interesting outcome is that however, even if education has a price, if it is admissible or it is accepted by the society, then knowledge still permits to reduce social inequalities. 
%Also, if the fully coupled model in which knowledge and social status influence each other, while inequalities in educational distribution are shown to increase,  the coupling maintains its reduction character with respect to social stratification, without any appreciable improvement of the social inequalities. 

\section*{Acknowledgment}
 This work has been written within the
activities of GNFM group  of INdAM (National Institute of
High Mathematics), and partially supported by  MIUR project ``Optimal mass
transportation, geometrical and functional inequalities with applications''.
The research was partially supported by
the Italian Ministry of Education, University and Research (MIUR): Dipartimenti
di Eccellenza Program (2018--2022) - Dept. of Mathematics ``F.Casorati'', University of Pavia. 
G.D. would like to thank the Italian Ministry of Instruction, University and Research (MIUR) to support this research with funds coming from PRIN Project 2017 (No. 2017KKJP4X entitled  ``Innovative numerical methods for evolutionary partial differential equations and applications'').

\end{document}